\documentclass[12pt]{article}

\usepackage[utf8]{inputenc}
\usepackage[english]{babel}
\usepackage{amssymb} 
\usepackage{amsmath}
\usepackage{amsthm}
\usepackage{latexsym}
\usepackage{graphicx}
\usepackage{bm}
\usepackage{bbm}
\usepackage{overpic}
\usepackage[normalem]{ulem}
\usepackage{url}
\usepackage{soul}
\usepackage{exscale}
\usepackage{amsfonts}
\usepackage[usenames,dvipsnames]{color} 
\usepackage{hyperref}
\usepackage{cleveref}
\usepackage{verbatim}
\usepackage[usenames,dvipsnames]{xcolor}
\usepackage{comment}
\usepackage{epsfig}
\usepackage{breqn}
\usepackage{enumitem}
\usepackage[draft]{todonotes}
\usepackage{subcaption}
\usepackage{float}
\usepackage[square,sort,comma,numbers]{natbib}
\usepackage{wrapfig}

\newtheorem{Definition}{Definition}[section]

\newtheorem{Remark}{Remark}[section]

\newtheorem{Theorem}{Theorem}[section]

\newtheorem{Proposition}{Proposition}[section]
\newtheorem{Assumption}{Assumption}[section]

\numberwithin{equation}{section}
\numberwithin{figure}{section}

\newcommand{\beas}{\begin{eqnarray*}}
\newcommand{\eeas}{\end{eqnarray*}}
\newcommand{\bea}{\begin{eqnarray}}
\newcommand{\eea}{\end{eqnarray}}
\newcommand{\ben}{\begin{enumerate}}
\newcommand{\een}{\end{enumerate}}

\newcommand{\bi}{\begin{itemize}}
\newcommand{\ei}{\end{itemize}}
\newcommand{\beq}{\begin{equation}}
\newcommand{\eeq}{\end{equation}}
\newcommand{\bv}{\begin{verbatim}}

  

\makeatletter
\newcommand*{\centerfloat}{%
  \parindent \z@
  \leftskip \z@ \@plus 1fil \@minus \textwidth
  \rightskip\leftskip
  \parfillskip \z@skip}
\makeatother

\begin{document}
\title{\bf Deep Hedging under Rough Volatility}

\author{Blanka Horvath\\ King's College London and The Alan Turing Institute,\\ {\tt blanka.horvath@kcl.ac.uk},\\ Josef Teichmann\\ ETH Z\"{u}rich\\ {\tt josef.teichmann@math.ethz.ch},\\ \ \v{Z}an \v{Z}uri\v{c}\\ Imperial College London\\ {\tt z.zuric19@imperial.ac.uk}, \\ }
\maketitle

\begin{abstract}
We investigate the performance of the Deep Hedging framework under training paths beyond the (finite dimensional) Markovian setup. In particular we analyse the hedging performance of the original architecture under rough volatility models with view to existing theoretical results for those. Furthermore, we suggest parsimonious but suitable network architectures capable of capturing the non-Markoviantity of time-series. Secondly, we analyse the hedging behaviour in these models in terms of P\&L distributions and draw comparisons to jump diffusion models if the the rebalancing frequency is realistically small.
\end{abstract}

\pagebreak

\section{Introduction}

Deep learning has undoubtedly had a major impact on financial modelling in the past years and has pushed the boundaries further of the challenges that can be tackled: Not only can existing problems be solved faster and more efficiently \citep{Hernandez2016,Horvath2019,Bayer2019,Liu2019,Ruf2019,Benth2020,Cuchiero2020,Gierjatowicz2020}, but deep learning also allows us to derive (approximative) solutions to optimisations problems \citep{Buehler2019}, where classical solutions had so far been limited in scope and generality. Additionally these approaches are fundamentally data driven, which makes them particularly attractive from business perspectives.

It comes as no surprise that the more similar (or ``representative") the data presented to the network in the training phase is to the (unseen) test data, which the network is later applied to, the better is the performance of the hedging network on real data (in terms of P{\&}L). It is also unsurprising that, as markets shift sufficiently far away from a presented regime into new, previously unseen territories, the hedging networks may have to be retrained to adapt to the new environment.

In the current paper we go a step further than just presenting an ad hoc well chosen market simulator (see \cite{HenryLabordere2019, Wiese2019, Wiese2020, Kondratyev2020, Buehler2020, Buehler2020a, Cuchiero2020a, Xu2020}): we investigate a situation where the relevant test data is \emph{structurally} so different from the original modelling setup that it calls for an adjustment of the model architecture itself: in a well-controlled synthetic data environment we study the behaviour of the hedging engine as relevant properties of the data change.

More specifically, we use synthetic data generated from a rough volatility model with varying levels of the Hurst parameter. In its initial setup we set the Hurst parameter to $H=1/2$, which reflects a classical (finite dimensional) Markovian case, which is well-aligned with the majority of the most popular classical financial market models, such as, e.g., the Heston model, which the initial version of the deep hedging results were demonstrated on. We then gradually alter the level of the Hurst parameter to (rough) levels around $H\approx0.1$, which more realistically reflects market reality as observed in \citep{Alos2007, Fukasawa2010, Gatheral2018, Bolko2020, Livieri2018} thereby introducing a non-Markovian \emph{memory} into the volatility process.

Since rough volatility models are known to reflect the reality of financial markets (as well as the stylised statistical facts) better than classical, finite-dimensional Markovian models do, our findings also give an indication how a naive application of model architectures to real data could lead to substantial errors. With this our study allows us to make a number of interesting observations for deep hedging and the data that it is applied to: apart from drawing parallels between discretely observed rough volatility models and jump processes, our findings highlight the need to rethink (or carefully design) risk management frameworks of deep learning models as significant structural shifts in the data occur.

The paper is organised as follows: Section \ref{sec:setup} recalls the setup of the original deep hedging framework used in \citep{Buehler2019}. Section \ref{sec:network} gives a brief reminder on hedging under rough volatility models and compares the performance of (feed-forward) hedging network on a rough Bergomi model compared to a theoretically derived model hedge.
In Sections \ref{sec:deepvsmodelperformance} and \ref{sec:implicationrchitecture} we draw conclusions with respect to the model architecture and in Section \ref{sec:proposedarchitecture} we propose a new architecture that is better suited to the data. Section \ref{sec:performance} lays out the hedging under the new architecture and draws conclusions to existing literature which outlines some parallels between (continuous) rough volatility models and jump processes in this setting, while Section \ref{sec:conclusion} summarizes our conclusions.

\section{Setup and Notation}\label{sec:setup}

We adopt the setting in \citep{Buehler2019} and consider a discrete finite-time financial market with  time horizon $[0,T]$ for some $T\in(0,\infty)$ and a finite number of trading dates $0 = t_0 < t_1 < \dots < t_n = T$, $n\in\mathbb{N}$. We work on a discrete probability space $(\Omega, \mathcal{F}, \mathbb{P})$, with $\Omega = \left\lbrace \omega_1, \dots , \omega_N \right\rbrace$ and a probability measure $\mathbb{P}$ for which $\mathbb{P} \left[ \lbrace \omega_i \rbrace \right] > 0$ for all $i\in\{1,\dots,N\}$ and $N\in\mathbb{N}$. Additionally, we fix the notation $\mathcal{X} := \{X: \Omega \rightarrow \mathbb{R}\}$ for the set of all $\mathbb{R}$-valued random variables on $\Omega$.\\
The filtration $\mathbb{F} = \left( \mathcal{F}_k \right)_{k=0,\dots,n}$ is generated by the $\mathbb{R}^r$-valued information process $(I_k)_{k=0,\dots,n}$ for some $n,r\in\mathbb{N}$. For any $k \in \{0,\dots,n\}$, the variable $I_k$ denotes all available new market information at time $t_k$ and $\mathcal{F}_k$ represents all available market information up to time $t_k$.\\ 
The market contains $d\in\mathbb{N}$ financial instruments which can be used for hedging, with mid-prices given by an $\mathbb{R}^d$-valued $\mathbb{F}$-adapted stochastic process $S = (S_k)_{k=0,\dots,n}$. In order to hedge a claim $Z: \Omega \rightarrow \mathbb{R}$ we may trade in $S$ according to $\mathbb{R}^d$-valued $\mathbb{F}$-adapted processes (strategies), which we denote by $\delta := (\delta_k)_{k=1,\dots,n}$, where $\delta_k=(\delta_k^1 ,\ldots ,\delta_k^d)$. Here, $\delta_k^i$ denotes the agent’s holdings of the $i$-th asset at time $t_k$. We denote the initial cash injected at time $t_0$ by $p_0>0$. \\
Furthermore, in order to allow for proportional trading costs, for every time $t_k$ and change in position $s\in \mathbb{R}^d$ we consider costs $c_k: s \mapsto c\in [0,\infty)$, where $c_k$ is $\mathcal{F}_k$-adapted, upper-semi continuous and for which $c_k(0)=0$ for all $k\in\{0,\dots,n\}$. The total costs up to time $T$, when trading according to a trading strategy $\delta$ are denoted by $C_T(\delta) := \sum_{k=0}^n c_k s_{k-1}(\delta_k - \delta_{k-1})$. Finally, we denote by $\mathcal{H}$ a set of all trading strategies.

\noindent We consider optimality of hedging under \emph{convex risk measures} as in \citep{Buehler2019, Xu2005} and \citep{IAR09}.  For a reminder on convex risk measures see e.g. \citep{FS16}.
Now let $\rho:\mathcal{X}\rightarrow\mathbb{R}$ be a cash invariant convex risk measure on the set $\mathcal{X}$. As in \citep{Buehler2019}, we consider for random variables $X\in\mathcal{X}$ the original optimization problem
\begin{equation}\label{eq:Optimal_riskmeasure}
-\pi (X) := \inf_{\delta\in\mathcal{H}}\rho(- X + (\delta\cdot S)_T - C_T(\delta)).
\end{equation}
An \emph{optimal hedging strategy} for $X$ is a minimizer $\delta \in \mathcal{H}$ of \eqref{eq:Optimal}, where the premium is $\pi(X) $.
~\\~\\
In case of no trading costs there is an alternative view point, which will be taken in this paper: consider an equivalent pricing measure $\mathbb Q$ of our financial market, then we can also minimize the variance (with respect to the pricing measure $\mathbb Q$)
\begin{equation}\label{eq:Optimal}
\inf_{\delta\in\mathcal{H}}\mathbb E\left[\left(X - \left(\delta\cdot S\right)_T - p_0\right)^2\right],
\end{equation}
where $p_0$ denotes the expectation of $X$ with respect to $\mathbb Q$, i.e.~the risk neutral price. In other words: the price of the quadratic hedging loss (payoff) should be minimal.

In the rest of this paper, the above optimisation \eqref{eq:Optimal} problem and--corresponding optimisers-- are considered in terms of their numerical approximation in the framework of \emph{hedging in a neural network setting} as formulated in \citep{Buehler2019}. In the remainder of this section we recall the notation and definitions to formulate this approximation property and the conditions that ensure its validity.
\begin{Definition}[Set of Neural Networks with a fixed activation function]
We denote by $\mathcal{NN}_{\infty, d_0, d_1}^\sigma$ the set of all NNs mapping from $\mathbb{R}^{d_0}\rightarrow \mathbb{R}^{d_1}$ with a fixed activation function $\sigma$. The set $\{\mathcal{NN}_{M,d_0,d_1}^\sigma\}_{M\in\mathbb{N}}$ is then a sequence of subsets in $\mathcal{NN}_{\infty, d_0, d_1}$ for which $\mathcal{NN}_{M,d_0,d_1}^\sigma = \{F^{\theta}: \theta \in \Theta_{M,d_0,d_1}\}$ with $\Theta_{M,d_0,d_1}\subset \mathbb{R}^q$ for some $q(M), M \in \mathbb{N}$.
\end{Definition}

\begin{Definition}\label{def:NN}
We call $\mathcal{H}_M\subset\mathcal{H}$ the set of unconstrained neural network trading strategies:
\begin{align}\label{eq:HM}
\begin{split}
\mathcal{H}_M &= \left\{(\delta_k)_{k=0,\dots,n-1}\in\mathcal{H}: \delta_k = F_k(I_0,\dots,\delta_{k-1}), F_k\in\mathcal{NN}_{M,r(k+1)+d,d}\right\}\\
&= \left\{(\delta_k)_{k=0,\dots,n-1}\in\mathcal{H}: \delta_k = F_k^{\theta_k}(I_0,\dots,\delta_{k-1}), \theta_k \in \Theta_{M,r(k+1)+d,d}\right\}
\end{split}
	\end{align}
\end{Definition}
We now replace the set $\mathcal{H}$ in \eqref{eq:Optimal} 
by the finite subset $\mathcal{H}_M\subset\mathcal{H}$. The optimisation problem then becomes
\begin{align}\label{eq:lambM}
\begin{split}
\pi^M(X) &:= \inf_{\delta\in\mathcal{H}_M} \rho \left(X + \left(\delta \cdot S\right)_T - C_T\delta\right) \\
&= \inf_{\theta\in\Theta_M} \rho \left(X + \left(\delta^\theta\cdot S\right)_T - C_T \delta^\theta\right)
\end{split}
\end{align}
where $\Theta_M = \prod_{k=0}^{n-1}\Theta_{M, r(k+1)+d, d}$ denote the network parameters from Definition \ref{def:NN}. With \eqref{eq:HM}, \eqref{eq:lambM} and Remark \ref{remark:markovianstructure}, the infinite-dimensional problem of finding an optimal hedging strategy is reduced to a finite-dimensional problem of finding the optimal NN parameters for the problem \eqref{eq:lambM}.
\begin{Remark}\label{remark:markovianstructure}
Note that in the above, we do not assume
that $S$ is an $(\mathbb{F},\mathbb{P})$-Markov process and that the contingent claim is of the form $Z:=g(S_T)$ for a payoff function $g:\mathbb{R}^d\rightarrow\mathbb{R}$. This would allow us to write the optimal strategy $\delta_k = f_k(I_k, \delta_{k-1})$ for some $f_k: \mathbb{R}^{r+d}\rightarrow\mathbb{R}^d$.
\end{Remark}
The next proposition recalls the central approximation property which states that the optimal trading strategy \eqref{eq:Optimal_riskmeasure} can be approximated by a semi-recurrent neural network of the form Figure \ref{fig:NNoriginal} in the sense that the functional $\pi^M(X)$ converges to $\pi(X)$,
 as $M$ becomes large.
\begin{Proposition}\label{prop:hornik}
Define $\mathcal{H}_M$ as in \eqref{eq:HM} and $\pi^M$ as in \eqref{eq:lambM}. Then for any $X\in\mathcal{X}$
\[
\lim_{M\rightarrow\infty} \pi^M(X) = \pi(X),
\]
where $\pi(X)$ denotes the optimal solution of the original optimisaton problem \eqref{eq:Optimal_riskmeasure}.
\end{Proposition}

\begin{Remark}
Of course there is a completely analogous formulation of this proposition for the optimal trading strategy \eqref{eq:Optimal}.
\end{Remark}

\begin{figure}[!hbt]
\centering
\includegraphics[width=1.\textwidth]{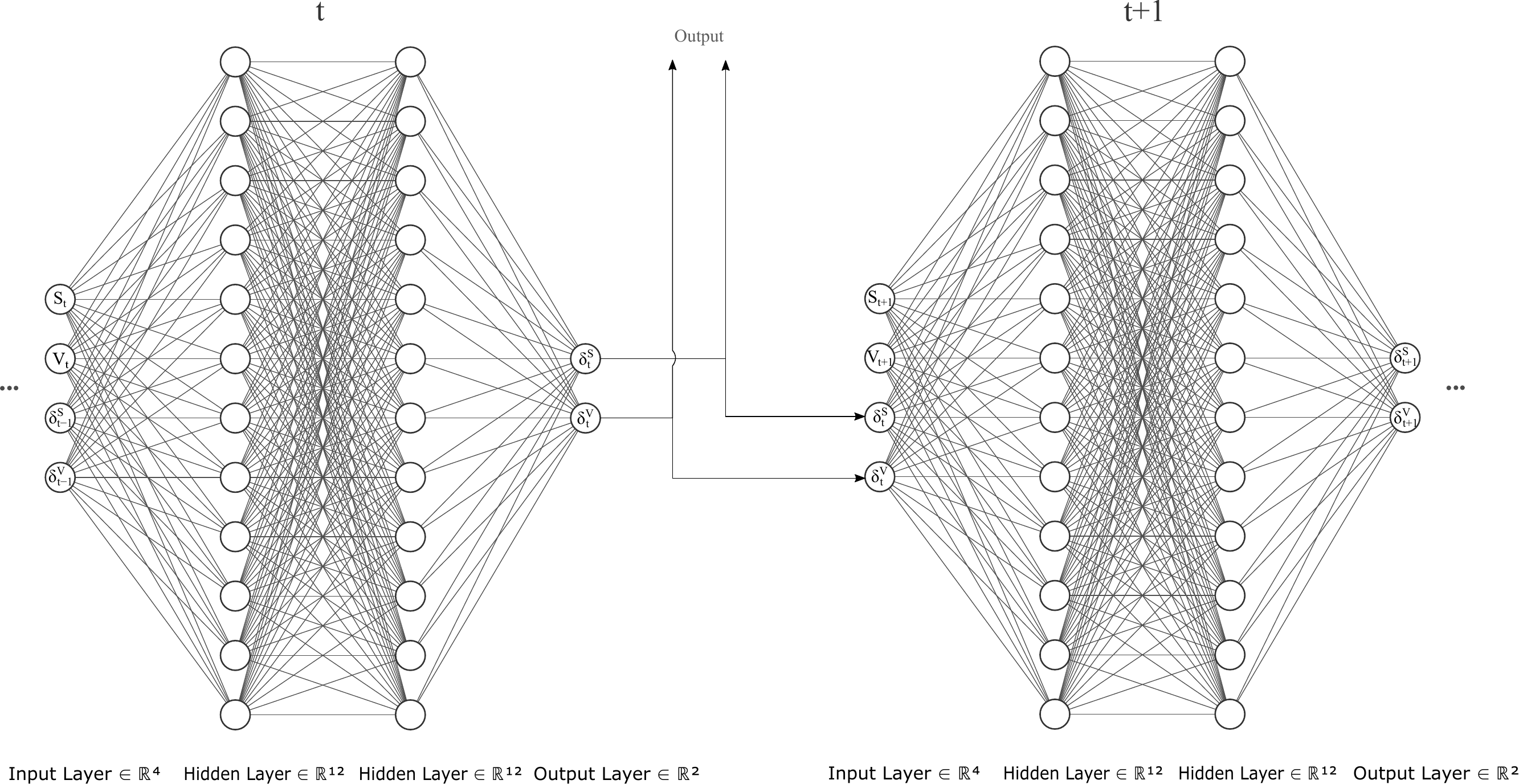}
\caption{Original Network Archiecture}\label{fig:NNoriginal}
\end{figure}
In \citep{Buehler2019} this approximation property is demonstrated for Black-Scholes and Heston models both in their original form and in variants including market frictions such as transaction costs. These results demonstrated how deep hedging which allows us to take a leap beyond classical results in scenarios where the Markovian structure is preserved.

A natural question to ask is, how the approximation property of the neural network is affected if the assumption of Markovian structure of the underlying process is no longer satisfied.
Rough Volatility models \citep{Alos2007, Fukasawa2010, Gatheral2018, Bayer2015} represent such a class of non-Markovian models. It is also well-established in a series of recent articles including the aforementioned works, that rough volatility dynamics are superior to standard standard Markovian models (such as Black-Scholes and Heston) in terms of reflecting market reality and also that rough volatility models are superior to a number of  in terms of allowing close fits to market data. 

By taking hedging behaviour under rough volatility models under the loop we gain insight into non-Markovian aspects of markets in a controlled numerical setting: Varying the Hurst parameter $H \in (0,1)$ of the process (see \cite{Gatheral2018}), which governs the deviation from the Markovian setting in a general fractional (or rough) volatility framework, enables us to control for the influence of the Markovianity assumption on the hedging performace of the deep neural network.
Therefore, in this work we investigate the effect of the loss of Markovianity property of the underlying stochastic process, by considering market dynamics that are governed in a rough volatility setting.  With this in mind, by applying the original feedforward network architecture to a more realistic model class (represented by rough volatility models) we in particular demonstrate how the the choice of the network architecture may affect the performance of Deep Hedging framework could potentially break down on real life data.
We also note in passing that the approach we take can be applied as a simple routine sanity check for model governance of deep learning models on real data:
\begin{itemize}
\item Take a well-understood model class that generalises the modelling to more realistic market scenarios, but where the generalisation no longer satisfy assumptions made in the original architecture.
\item Test the robustness of the method if the assumption is violated by controlling for the error as the deviation from the assumption increases.
\item Modify the network architecture accordingly  if necessary.
\end{itemize}

\section{Hedging and Network Architectures for Rough Volatility}\label{sec:network}
\subsection{Hedging under Rough Volatility}
Let us now consider the problem of hedging under rough volatility models in general. For this we
consider now a continuous filtration\footnote{For the numerical implementation of the resulting strategies that we consider in the following sections, we naturally consider again the discrete filtration introduced above in Section \ref{sec:setup}.} $\{\mathcal{F}_t\}_{0 \leq t \leq T}$. We know that for a Markovian process of the form 
\[
\tilde{X}_t = x + \int_0^t b(r, X_r)d r + \int_0^t \sigma(r, X_r)d W_r,
\]
where $b$ and $\sigma$ satisfy suitable conditions, the price of a contingent claim $\tilde{Z}_t := \mathbb{E}[g(\tilde{X}_T) \vert \mathcal{F}_t]$ can be written as
\[
\tilde{Z}_t = u(t, {\tilde{X}}_t),
\]
where $u$ solves a parabolic PDE by Feynman-Kac formula \citep{Kac1949}. However, it was shown in \citep{Bayer2015} that Rough volatility models are not finite dimensional Markovian and we therefore have to consider a more general process $X$ and assume it to be a solution to the $d$-dimensional Volterra SDE:
\begin{equation}\label{eq:volterraframework}
X_t = x + \int_0^t b(t; r, X_.) d r + \int_0^t \sigma (t; r, X_.) d W_r, \qquad t\in[0,T],
\end{equation}
where $W$ is a $m$-dimensional standard Brownian motion, $b\in\mathbb{R}^d$ and $\sigma\in\mathbb{R}^{m\times d}$. Both are adapted in a sense that for $\varphi = b, \sigma$ it holds $\varphi(t; r, X_.) = \varphi (t; r, X_{r \land .})$.

In this general non-Markovian framework, the contingent claim in the form $Z_t := \mathbb{E}[g(X_T) \vert \mathcal{F}_t]$ will depend on the entire history of the process $X := (X_t)_{t\geq 0}$ up to time $t$ and not just on the value of the process at that time  i.e.
\[
Z_t = u(t, X_{[0,t]}) \quad \text{ with notation } \quad X_{[0,t]} := \left\lbrace X_r \right\rbrace_{r\in[0,t]},
\]
where $u$ this time solves a Path dependent PDE (PPDE). The setting where $X$ is a \emph{semi-martingale} has already been explored in e.g. \citep{Dupire2019, Cont2013}. Be that as it may, we know that fBm is \emph{not} a semi-martingale in general and as a consequence the volatility process is not a semi-martingale. Viens and Zhang \cite{Viens2019} are able to cast the problem back in to the semi-martingale framework by rewriting $X_t$ as a orthogonal decomposition to an auxiliary process $\Theta_t$ and a process $I_t$, which is independent of the filtration
\begin{align}
\begin{split}
X_s &= x + \int_0^t b(s; r, X_.) d r + \int_0^t \sigma (s; r, X_.) d W_r \\ & \qquad + \int_t^s b(s; r, X_.) d r + \int_t^s \sigma (s; r, X_.) d W_r
\end{split}
\\
&:= x + \Theta_s^t + I_s^t
\end{align}
for $0\leq t \leq s$. By exploiting the semi-martingale property of $\Theta$, they go on to show that the contingent claim can be expressed as a solution of a PPDE
\begin{equation}
Z_t = u(t, X_{[0,t)} \otimes_{t} \Theta_{[t,T]}^t),
\end{equation}
where $\otimes_{t}$ denotes concatenation at time $t$. Moreover, they develop an Itô-type formula for a general non-Markovian process $X_t$ from \eqref{eq:volterraframework}, which we present in the Appendix. 
\subsection{The rough Bergomi model (rBergomi)}
As an example we consider the rBergomi with a constant initial forward variance curve $\xi_0(t)=V_0$:
\begin{subequations}
\begin{align}
S_t &= S_0 + \int_0^t S_r \sqrt{V_r} \left[ \sqrt{1-\rho^2} d B_r + \rho d W_r \right] \label{eq:rmergomis} \\
V_t &= V_0\mathcal{E}\left( \sqrt{2H}\nu \int_0^t (t-r)^{H-\frac{1}{2}} d W_r \right), \qquad  V_0 {=} v_0 {>} 0, \label{eq:rbergomiv}
\end{align}
\end{subequations}
The model fits into the affine structure of our Volterra SDE in \eqref{eq:volterraframework}, after a simple log-transformation of the volatility process. In this case we take our auxiliary process to be
\begin{equation}\label{eq:rbegomiauxiliary}
\Theta_s^t = \sqrt{2H}\nu\int_0^t (s-r)^{H-\frac{1}{2}} d W_r, \qquad t<s.
\end{equation}
It is easy to check that $\Theta^t_s$ is a true martingale for fixed $s$. The option price dynamics are obtained by using the Functional Itô formula in \eqref{eq:funcitoformula}. From this, the perfect hedge in terms of a forward variance $\hat{\Theta}_T^t$ with maturity $T$ and a stock $S_t$ follows:
\begin{align}\label{eq:rBergomihedge}
d Z_t = \partial_x u (t, S_t, \Theta_{[t,T]}^t) d S_t + \frac{(T-t)^{\frac{1}{2}-H}}{\hat{\Theta}_T^t} \left\langle \partial_\omega u(t, S_t, \Theta_{[t,T]}^t), a^t \right\rangle d \hat{\Theta}_T^t \\ \qquad \text{with } a^t_s = (s-t)^{H-\frac{1}{2}} \nonumber
\end{align}
The path-wise derivative in \eqref{eq:rBergomihedge} is the Gateaux derivative along the direction $a^t$. For more details and discretization of the Gateaux derivative see Appendix \ref{apx:pathderivatives}.

\subsection{Performance of the deep hedging (with the original feedforward architecture) scheme compared to the model hedge under rBergomi}\label{sec:deepvsmodelperformance}

We choose to hedge a plain vanilla call option $Z_T:=\max(S_T-K, 0)$ with $K=100$ and a monthly maturity $T = 30/365$. The hedging portfolio consists of a stock $S$ with $S_0=100$ and a forward variance with maturity $T_{\text{Fwd}}=45/365$ and is rebalanced daily. For the rBergomi model forward variance is equal to
\begin{equation}
\hat{\Theta}_{T_{\text{Fwd}}}^t := \mathbb{E}_{\mathbb{Q}}\left[ \int_0^{T_{\text{Fwd}}} V_s d s \middle\vert \mathcal{F}_t \right] = V_0 \exp \left[ \Theta_{T_{\text{Fwd}}}^t + \frac{1}{2}\nu^2 \left[ ({T_{\text{Fwd}}}-t)^{2H} - T_{\text{Fwd}}^{2H} \right] \right],
\end{equation}
{\sloppy with $\Theta_{T_{\text{Fwd}}}^t$ defined as in \ref{eq:rbegomiauxiliary}. Applying classical Itô's Lemma to $\hat{\Theta}_{T_{\text{Fwd}}}^t = \hat{\Theta}_{T_{\text{Fwd}}}^t(t, \Theta_{T_{\text{Fwd}}}^t)$ yields the dynamics of the forward variance under the rough Bergomi}
\begin{equation}
d \hat{\Theta}_{T_{\text{Fwd}}}^t = \hat{\Theta}_{T_{\text{Fwd}}}^t \sqrt{2H}\nu (T_{\text{Fwd}}-t)^{H-\frac{1}{2}} d W_t,
\end{equation}
which is well defined for $t\in[0,T_{\text{Fwd}})$. Therefore, choosing the maturity of the forward variance to be longer than the option maturity allows us to avoid the singularity as $t\rightarrow T$. In practice this would correspond to hedging with a forward variance with a slightly longer maturity than that of the option. 

For the simulation of the forward variance we used the Euler-Mayurama method, whereas paths of the volatility process were simulated with the ``turbo-charged'' version of the hybrid scheme proposed in \citep{Bennedsen2017, McCrickerd2018}. The parameters were chosen such that they describe a typical market scenario with a flat forward variance: $\xi_0 = 0.235\times 0.235$, $\nu = 1.9$ and $\rho = -0.7$. We were particularly interested in the dependence of the hedging loss on the Hurst parameter. Finally quadratic loss function was chosen and the minimizing objective was therefore
\[
\pi(-Z) = \inf_{\delta^\theta\in\mathcal{H}^M}\mathbb{E}\left[ (-Z + p_0 + (\delta^\theta\cdot S)_T)^2 \right]
\]
where price $p_0$ was obtained with a Monte-Carlo simulation (e.g. for $H=0.10$, $p_0=2.39$).

\begin{table}[h]
\centering
\begin{tabular}{ccc}
\hline
&  \multicolumn{2}{c}{Quadratic hedging loss} \\
\hline
$H$         & Model hedge               & Deep hedge \\ \hline \hline
\multicolumn{1}{c|}{$0.10$} & \multicolumn{1}{c|}{1.45} & 1.16 (*1.12)       \\
\multicolumn{1}{c|}{$0.20$} & \multicolumn{1}{c|}{0.52} & 0.67     \\
\multicolumn{1}{c|}{$0.30$} & \multicolumn{1}{c|}{0.34} & 0.46      \\
\multicolumn{1}{c|}{$0.40$} & \multicolumn{1}{c|}{0.24} & 0.36       \\
& & {\tiny *-on 200 epochs}
\end{tabular}
\caption{Comparison of the quadratic loss between model and deep hedges trained 75 epochs for different $H$.}\label{table:bergvsdeepH}
\end{table}

Next we implement the perfect hedge from \eqref{eq:rBergomihedge} the details of the discretization of the Gateaux derivative are presented in Appendix \ref{apx:discretegateux}. For evaluation of the option price, we once again use Monte-Carlo, this time with generating parameters. In practice we would calibrate the parameters to the market data. Perfect hedge was implemented on the sample of $10^3$ different paths for the same parameters as in the deep hedging case. The results of both hedges under quadratic loss for different Hurst parameters are shown in Table \ref{table:bergvsdeepH}.
We also take a closer look of the P\&L distributions of the deep hedge as well as the model hedge for $H=0.10$ in Figure \ref{fig:rBergomi01-rBergomi_vs_deep75_subsample10k}. Curiously enough, the distributions are very similar to each other. The deep hedge seems to have slightly thinner tails, which is interesting, considering the semi-recurrent architecture makes a strong assumption of Markovianity of the underlying process.

\begin{figure}[hbt!]
\includegraphics[width=1.\linewidth]{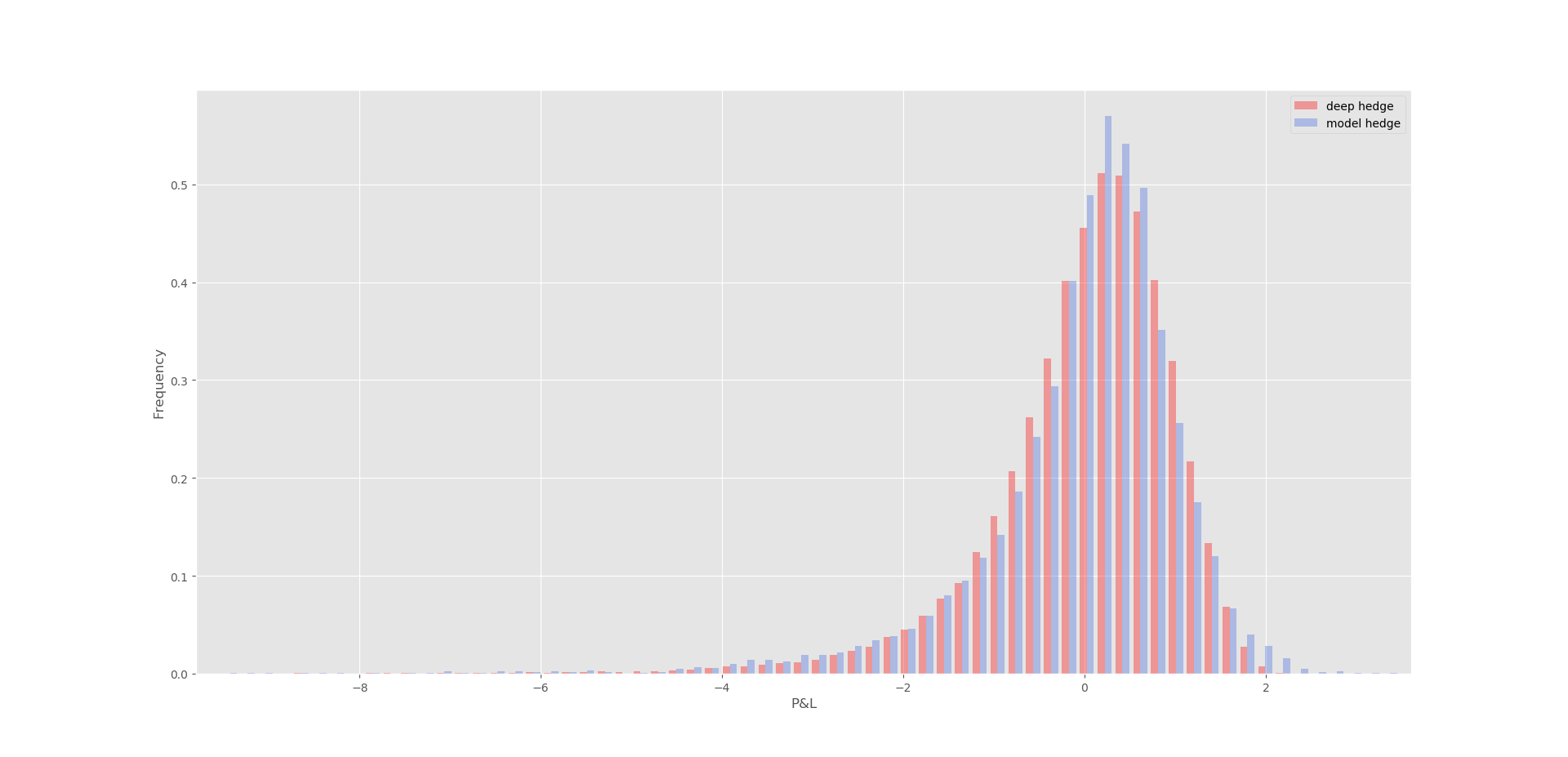}
\caption{rBergomi model hedge (\textit{blue}) compared to the deep hedge (\textit{red}) trained on $75$ epochs on rBergomi paths with $H=0.10$. Note that the option price is only $p_0=2.39$ and that such a hedge can result in a \emph{substantial} loss.}\label{fig:rBergomi01-rBergomi_vs_deep75_subsample10k}
\end{figure}
Indicators that the assumption of finite dimensional Markovianity is violated might be the heavy left tail of the P\&L distribution as well as relatively high hedging losses. This prompted us to question the semi-recurrent architecture and devise a way to relax the Markov assumption on the underlying. Note that the heavy tails of these distribution may also imply a link to jump diffusion models. We expand on this in Section \ref{sec:relationtolit}.
\subsection{Implications on the network architecture}\label{sec:implicationrchitecture}

As discussed before, in \citep{Buehler2019} authors heavily rely on Remark \ref{remark:markovianstructure}, where they use the Markov property of the underlying process in order to write the trading strategy at time $t_k$ as a function of the information process at $t_k$ and trading strategy in the previous time step $k-1$. Of course, in the case of rough volatility models one would have to include the entire history of the information process up to $t_k$ in order to get the hedge at that time. However, this would result in numerically infeasible scheme. To illustrate this, take for example a single vanilla call option with maturity $T=30/365$, where we hedge daily under say the rough Bergomi model. In the $30$-th time step the number of input nodes of the NN cell $F_{30}^\theta$ would be $30\cdot 2 + 2 = 62$ or if we hedged twice a day $30\cdot 2\cdot 2 + 2 = 122$. Obviously this scheme quickly becomes very computationally expensive even for a single option with a short maturity.

The fBm in \eqref{eq:rbergomiv} can be written as a linear functional of a Markov process, albeit an infinite dimensional one. Therefore, if the original Markovian-based architecture can be applied to this setting, we would expect to recover the Hurst parameter also from a Markovian-based sampling procedure, justifying the continued use of the original  feed forward architecture. This however is not  the case:
It is known that fBm in \eqref{eq:rbergomiv} can be rewritten as an infinite dimensional Markov process in the following way. Take the Riemann–Liouville representation of \textup{fBm}:
\[
B^H_{t} := \frac{1}{\Gamma (H + \frac{1}{2})}\left( \int_0^t (t-s)^{H - \frac{1}{2}} d W_s \right),
\]
where $W$ is a standard Brownian motion. Using the fact that for $\alpha\in (0,1)$ and fixed $x\in [0,\infty)$:
\begin{equation}
\frac{(t-s)^{\alpha-1}}{\Gamma (\alpha)} = \int_0^\infty e^{-(t-s)x}\mu( d x), \quad \text{with} \quad \mu( d x) = \frac{d x}{x^\alpha \Gamma (\alpha) \Gamma (1-\alpha)}
\end{equation}
we obtain by the Fubini Theorem
\begin{align*}
B^H_{t} &= \int_0^t \int_0^\infty e^{-(t-s)x} \mu(d x) d W_s \\
&= \int_0^\infty \int_0^t e^{-(t-s)x} d W_s \mu(d x)\\
&= \int_0^\infty Y_t^x \mu(d x)
\end{align*}
with $Y_t^x = \int_0^t e^{-(t-s)x} d W_s$. Observe that for a fixed $x\in[0,\infty)$, $(Y_t^x)_{t\geq 0}$ is an \textit{Ornstein-Uhlenbeck process} with mean reversion zero and mean reversion speed $x$ i.e. Gaussian semi-martingale Markov process solution with the dynamics of
\[
d Y_t^x = -x Y_t^x d t + d W_t.
\]
Therefore, we have shown that $B^H$ is a linear functional of the infinite dimensional Markov process. Being able to simulate from $Y_t^x$ would mean that we can still use the architecture in Figure \ref{fig:NNoriginal}, even for a rough processes.
Numerical simulation scheme for such a process is presented in \citep{Carmona1998}. Regrettably the estimated Hurst parameter\footnote{Several estimation procedures of the Hurst parameter were used see e.g. \citep{Matteo2005, Matteo2007}. Estimations of the paths simulated with the hybrid scheme \citep{Bennedsen2017, McCrickerd2018} were on the other hand in alignment with the input parameter.} from the generated time series stayed around $H\approx 0.5$, for any chosen input Hurst parameter to the simulation scheme. For a fixed time-step $\Delta t$ the scheme does not produce desired roughness, even if we used number of OU-terms well beyond what authors propose. We believe this is because scheme is only valid in the limit i.e. when the number of terms goes to infnity and $\Delta t\rightarrow 0$. Failure to recover the Hurst parameter together with the fact that the architecture does not allow for any path dependent contingent claims, encouraged us to change the Neural Network architecture itself.

\subsection{Proposed fully recurrent architecture}\label{sec:proposedarchitecture}
\begin{figure}[!hbt]
\centering
\includegraphics[width=1.\textwidth]{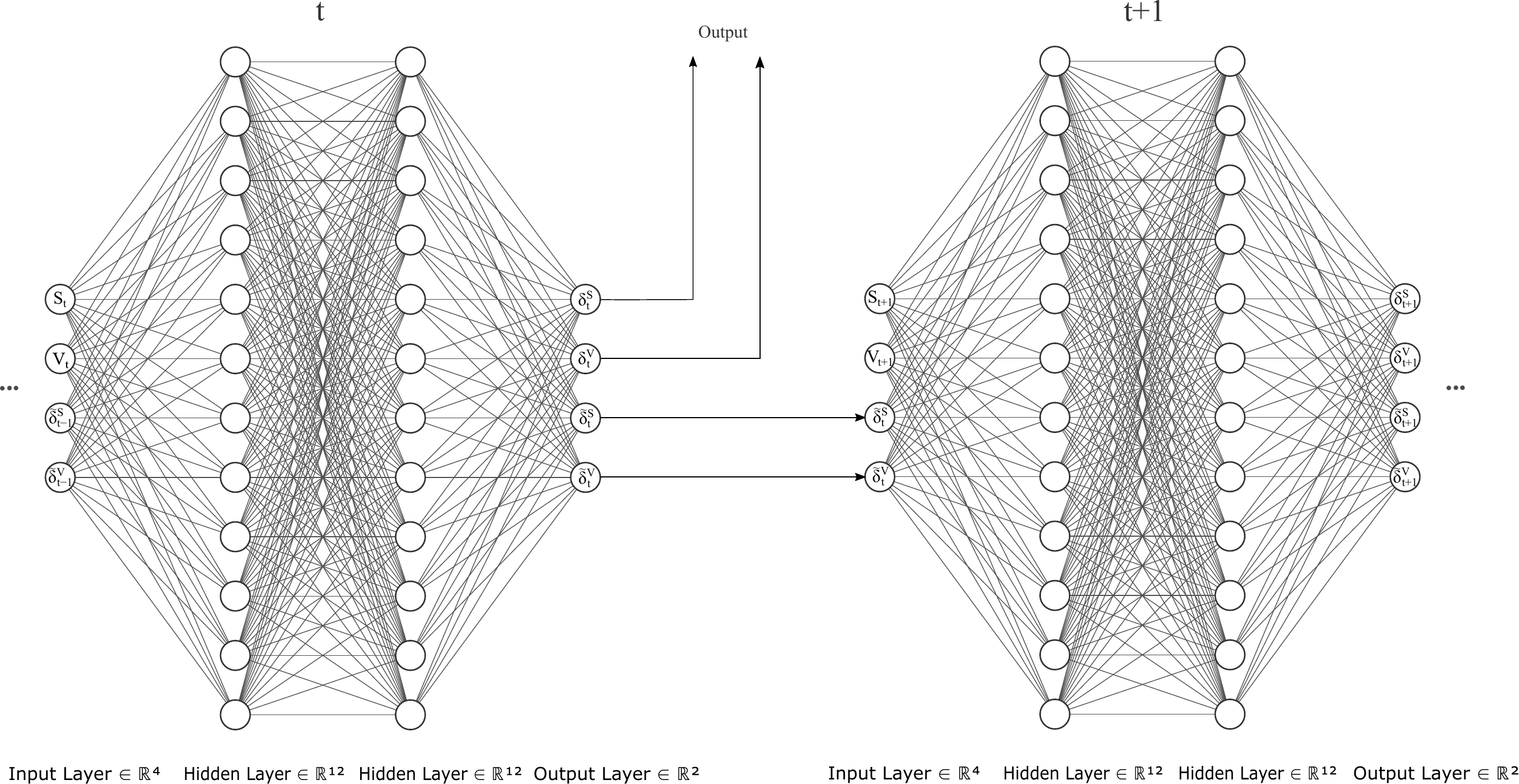}
\caption{Fully Recurrent Neural Network (fRNN) Architecture. The recurrent structure of this architecture is clearly visible as hidden states are passed on to the next cell at each time step.}\label{fig:NNproposed}
\end{figure}
By the above insights we hence modify the original architecture. In this section we suggest an alternative architecture and show that it is well-suited to the problem. When constructing a new architecture, we would like to change the semi-recurrent structure as little as possible for our purpose, since it seems to perform very well in the Markovian cases. However, in order to account for non-Markovianity we propose a completely recurrent structure.\footnote{Note that by completely recurrent we do not mean the same network is used at each time step, but that the hidden state is passed on to the cell in the next time step along with current portfolio positions.} To that end we now introduce a hidden state $\tilde{\delta}_k = (\tilde{\delta}^S_{k-1}, \tilde{\delta}^V_{k-1})$ with $\tilde{\delta}_0 = 0$, which is passed to the cell at time $t_k$ along the information process $I_k$. So instead of adding layers to each of the state transitions separately as in \citep{Pascanu2013}, we simply concatenate the input vector $I_k$ with the hidden state vector and feed it into a the neural network cell $F_k^{\theta}(\cdot)$:
\[
F_k^{\theta}\left(I_k \oplus \tilde{\delta}_k \right) = \delta_k \oplus \tilde{\delta}_k
\]
For the visual representation see Figure \ref{fig:NNproposed}. The \textit{output} is still a trading strategy $\delta_k = (\delta_k^S, \delta_k^V)$ and it is evaluated on the same objective function as before in case of quadratic hedging losses (without transaction costs):
\[
\mathcal{L}(\theta) := \mathbb{E}\left[ \left( - Z + p_0 + (\delta^\theta \cdot S)_T) \right)^2 \right],
\]
whereas the hidden state $\tilde{\delta}_k$ is passed forward to the next cell $F_{k+1}^\theta$. These states can take any value and are not restricted to having any meaningful financial representation as trading strategies do. We illustrate the fact that the fRNN architecture is truly recurrent by showing how hidden states are able to encode the relevant history of the information process. Let's say for example that the information process $I_k = (S_k^1, S_k^2)$ is simply the price of both hedging instruments. The strategies at time $t_k$ now do not depend on the asset holdings $\delta_{k-1}^x$, but on $\tilde{\delta}^x_{k-1}$ for $x\in\{S,V\}$:
\begin{align*}
\delta_k^x &:= \delta_k^x(S^1_k, S^2_k, \tilde{\delta}^S_{k-1}, \tilde{\delta}^V_{k-1}). 
\end{align*}
For some $\mathcal{F}_{k-1}$-measurable function $g_{k-1}$, it holds for the hidden states themselves that
\begin{align*}
\tilde{\delta}_{k-1}^x &= g^x_{k-1}(S^1_{k-1}, S^2_{k-1}, \tilde{\delta}^S_{k-2}, \tilde{\delta}^V_{k-2}). 
\end{align*}
Recursively the hidden states are implicitly dependent on the entire history
\begin{align*}
\tilde{\delta}_{k-1}^x &= g_{\mathcal{NN}}^x(S^1_{k-1}, S^2_{k-1}, S^1_{k-2}, S^2_{k-2},\dots, S^1_0, S^2_0, \tilde{\delta}_0), 
\end{align*}
where $g_{\mathcal{NN}}^x$ is again $\mathcal{F}_{k-1}$-measurable. Structuring the network this way, we are hoping that the hidden states at time $t_k$ will be able to encode the history of the information process $I_0,\dots,I_k$. More precisely, what we expect is that the network will learn itself the function $g_{\mathcal{NN}}^x: \mathbb{R}^{2k} \rightarrow \mathbb{R}$ for $x\in\{S,V\}$ and with that the path dependency inherit to the liability we are trying to hedge.
\begin{Remark}
We remark that in order to account for the history of the information process one could also write the trading strategy as
\[
\delta_k := \delta_k(I_k, \tilde{I}_{k-1}),
\]
where $\tilde{I}_{k-1}^n = \{I_i\}_{i=(k-1)-n}^{k-1}$ is the history of the information process with a window length of $n\in\{1,\dots,k-1\}$. However in this case, we would have to optimize the window length and would inevitably face an accuracy and computational efficiency trade-off. We would rather outsource this task to the neural network.
\end{Remark}
\begin{Remark}
While we do think LSTM architecture \citep{Hochreiter1997} would be more appropriate to capture the non-Markovian aspect of our process, we find that our architecture is adequate in that regard as well. Our architecture has the advantage of being tractable (we can still appeal to the Proposition \ref{prop:hornik}), all while being much simpler and easier to train.
\end{Remark}

\section{Hedging Performance and Hedging P\&L under the Rough Bergomi Model}\label{sec:performance}
\subsection{Deep hedge under Rough Bergomi}

Since the fRNN should perform just as well in Markovian case as the original one does, we first convinced ourselves that our architecture produces comparable results in the classical case. Quadratic losses as well as the training time for the Heston model were very similar for both\footnote{For Heston parameters $\alpha=1, b=0.04, \sigma = 0.8, V_0=0.04, S_0=100 \text{ and } \rho=-0.7$ the quadratic losses were $0.20$ under original architecture and $0.162$ under the fully recurrent one. Both training times were fairly similar as well.}. We were now ready to test it on the rough Bergomi model. We hedge the ATM call from Section \ref{sec:deepvsmodelperformance}, the parameters were again $\xi_0 = 0.235\times 0.235$, $\nu = 1.9$ and $\rho = -0.7$ and we investigate the dependence of the hedging loss on the Hurst parameter. The results are shown in Table \ref{table:deephedgevsH}. Again, the loss seems to exponentially decrease with increasing Hurst parameter and reaches quadratic losses comparable to classic stochastic volatility models at $H \gtrsim 0.5$.

\begin{table}[h]
\centering
\begin{tabular}{ccccccccc}
\hline
&  \multicolumn{8}{c}{Hurst parameter} \\
\hline
$H$  & $0.10$  & $0.20$ & $0.30$  & $0.40$ & $0.60$ & $0.70$ & $0.80$ & $0.90$ \\ \hline \hline
\multicolumn{1}{c|}{Quad. loss} & 0.834 (*0.628) & 0.376 & 0.263 & 0.244 & 0.204 & 0.206 & 0.197 & 0.191 \\
\multicolumn{7}{c}{} & \multicolumn{2}{c}{\tiny* - loss on 200 epochs}
\end{tabular}
\caption{Quadratic loss for different Hurst parameters. Run time on 75 epochs was approximately 2 hours for each parameter.}\label{table:deephedgevsH}
\end{table}

Comparing these results, with both the model hedge and the deep hedge from Section \ref{sec:deepvsmodelperformance} (see Table \ref{table:bergvsdeepH-proposed}), we notice the fRNN does indeed perform \emph{notably} better. By increasing number of epochs in the training phase from 75 to 200 the loss in the case of the deep hedge with original architecture does not improve, while the improvement with the proposed architecture is clearly visible. This indicates that while the semi-recurrent NN saturates at a given error, the new architecture keeps converging and improving. Since the training at 200 epochs was computationally costly (in terms of both memory and time) and since we have reached the model hedge's numbers at the higher end of $H$ range we did not keep increasing the number of epochs. But we expect that to keep improving as the number of epochs increases, which definitely indicates the second approaches suitability.

\begin{table}[h]
\centering
\begin{tabular}{cccc}
\hline
&  \multicolumn{3}{c}{Quadratic hedging loss} \\
\hline
$H$        & Model hedge               & Deep hedge                        & Deep hedge - fRNN \\ \hline
\multicolumn{1}{c|}{$0.10$} & \multicolumn{1}{c|}{1.45} & \multicolumn{1}{c|}{1.16 (*1.12)} & 0.83 (*0.63)          \\
\multicolumn{1}{c|}{$0.20$} & \multicolumn{1}{c|}{0.52} & \multicolumn{1}{c|}{0.67}         & 0.38                  \\
\multicolumn{1}{c|}{$0.30$} & \multicolumn{1}{c|}{0.34} & \multicolumn{1}{c|}{0.46}         & 0.26                  \\
\multicolumn{1}{c|}{$0.40$} & \multicolumn{1}{c|}{0.24} & \multicolumn{1}{c|}{0.36}         & 0.22                  \\
& & &  \multicolumn{1}{r}{\tiny *-on 200 epochs}
\end{tabular}
\caption{Comparison of the quadratic loss between model and deep hedges with \emph{fRNN architecture} trained on 75 epochs for different $H$.}\label{table:bergvsdeepH-proposed}
\end{table}

\begin{figure}[!hbt]
\centering
\includegraphics[width=0.7\textwidth]{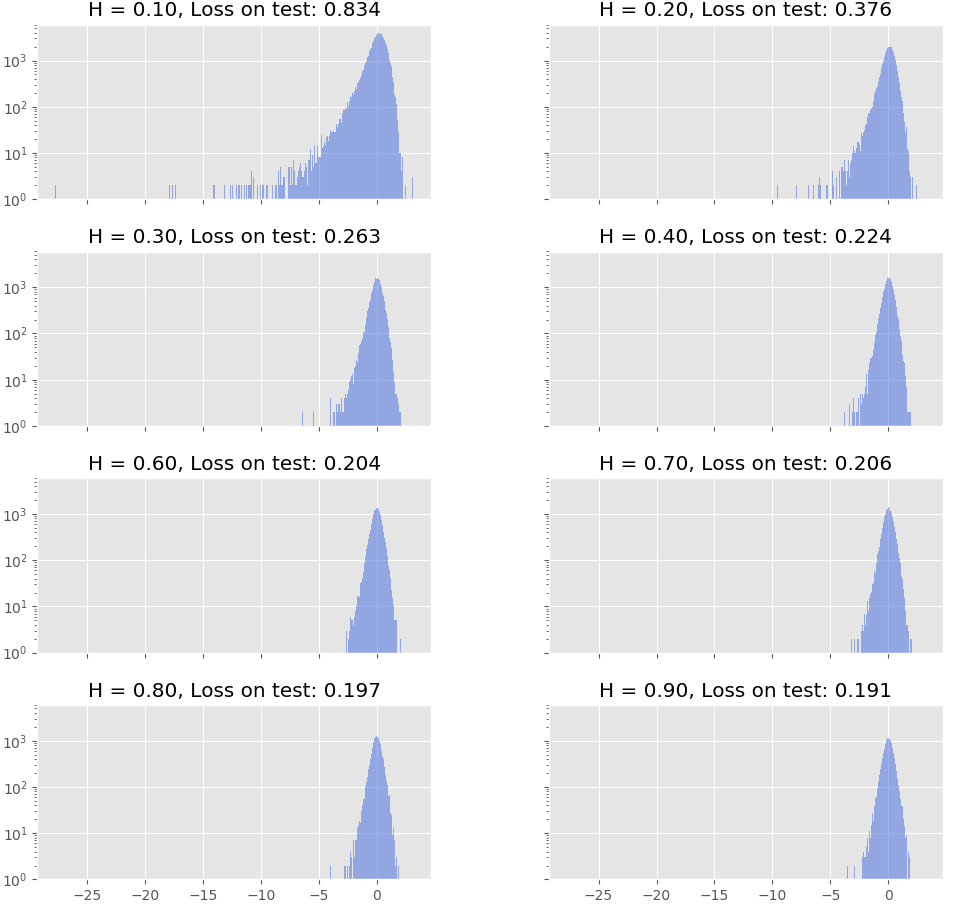}
\caption{Empirical P\&L distributions in log-scale for different Hurst parameters under the fRNN hedge. \textit{Loss on test} denotes the realized quadratic loss on the test set for a network trained on 75 epochs.}\label{fig:deephedgingPLs}
\end{figure}

Looking at the Figure \ref{fig:deephedgingPLs} it is particularly interesting that the P{\&}L distribution becomes increasingly left tailed with lower Hurst parameters. Even under the new architecture the distribution for $H=0.10$ is left-skewed with an extremely heavy left tail, where relative losses reached cca. $-1000\%$ in one of $10^5$ sample paths. What is even more compelling is that the sizeable losses occurred, when the discretized stock process \emph{jumped} by several thousand basis points during the hedging period. Example of such a path is shown in Figure \ref{fig:stock_jump}. Although jumps are not featured in the rough Bergomi model (the price process is a continuous martingale \citep{Gassiat2019}) the model clearly exhibits jump-like behaviour when \emph{discretized}. 

\begin{figure}[!hbt]
\centering
 \includegraphics[width=.5\textwidth, height=0.25\textheight, trim={2cm 2cm 2cm 2cm},clip]{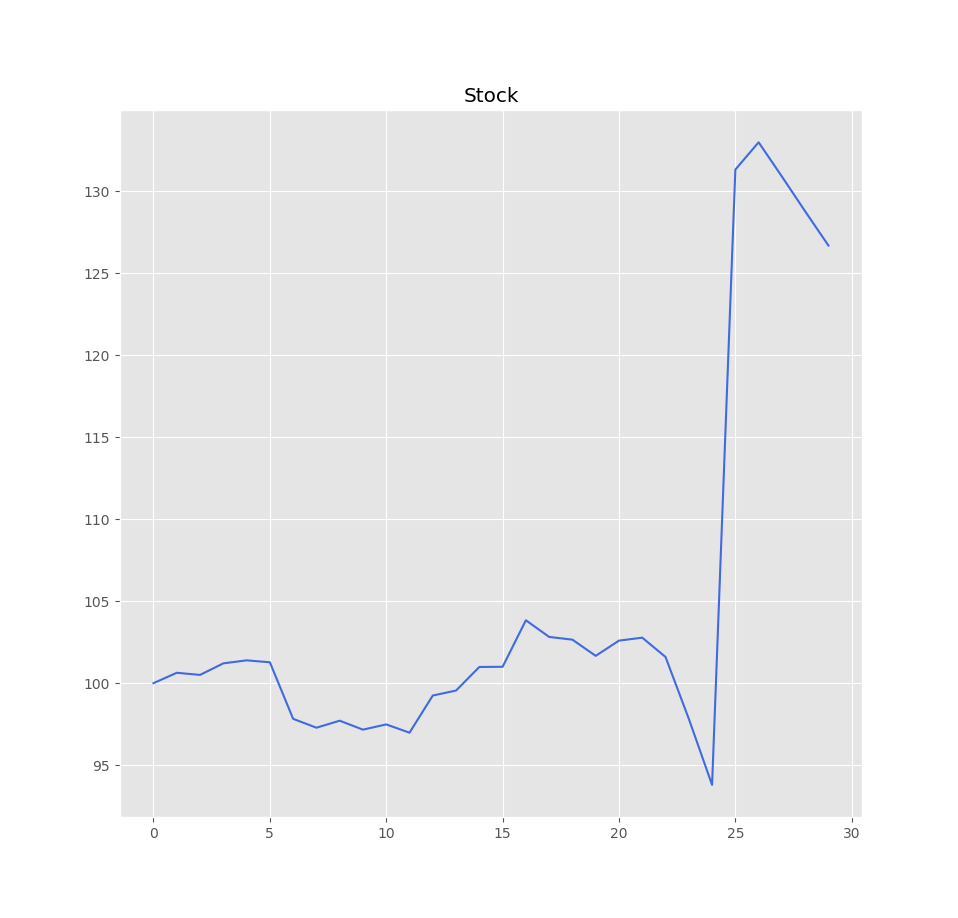}
\caption{Under the discretized rough Bergomi model the stock can jump by more than $\pm 30\%$ in a single time step. This stock path caused extreme loss of $-27.73$ seen in Figure \ref{fig:deephedgingPLs}.}\label{fig:stock_jump}
\end{figure}

Naturally, for $H=0.10$, where this effect was the most noticeable, we tried increasing the training, test and validation set sizes, as well as number of epochs to $200$. Doing this we managed to decrease the realized loss to $0.628$. The performance was notably better compared to $0.834$ on smaller set sizes, but still far from the loss of $0.162$ we obtained under the Heston model. We investigated settings more epochs, bigger training sizes, different architectures, however the realized test loss did not improve.

As it can be seen in Figure \ref{fig:bergvsdeepH} model hedge loss distribution exhibits very similar behaviour as the deep hedge distribution. Higher losses of the model hedge can be explained by the slightly fatter tail in comparison to the fully recurrent hedge. We remark this behaviour is somewhat understandable, since re-hedging is done daily and the hedging frequency is far from being a valid approximation for a continuous hedge. In the next section we thus implement hedges at different frequencies to see, whether the Hölder regularity of the underlying process is problematic only for the deep hedging procedure or is the heavy left-tailed P\&L distribution a general phenomena, when hedging under a discretized rough model.

\begin{figure}[!hbt]
\begin{subfigure}{.5\textwidth}
\includegraphics[width=1.\linewidth]{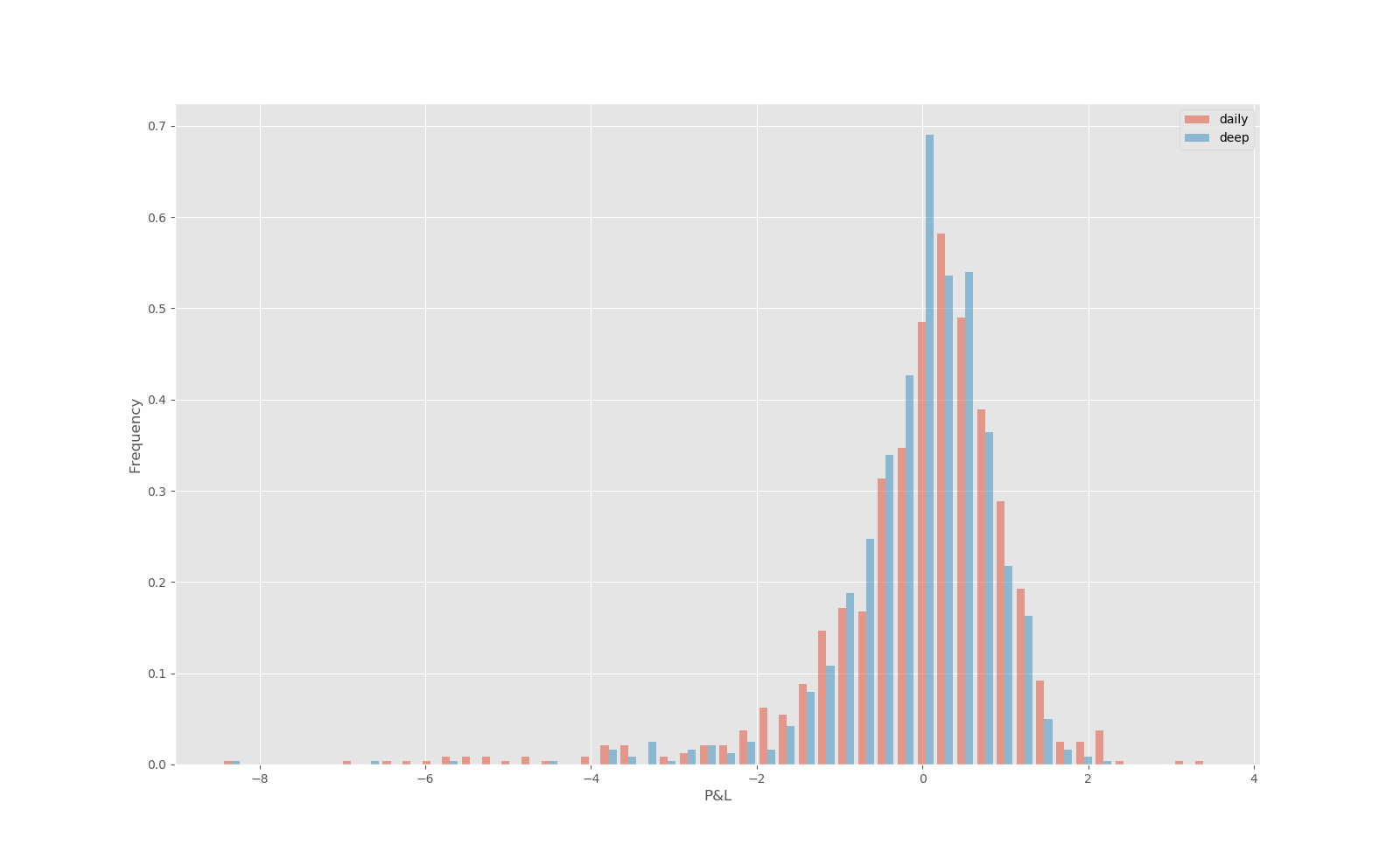}
\caption{H=0.10}
\end{subfigure}%
\begin{subfigure}{.5\textwidth}
\includegraphics[width=1.\linewidth]{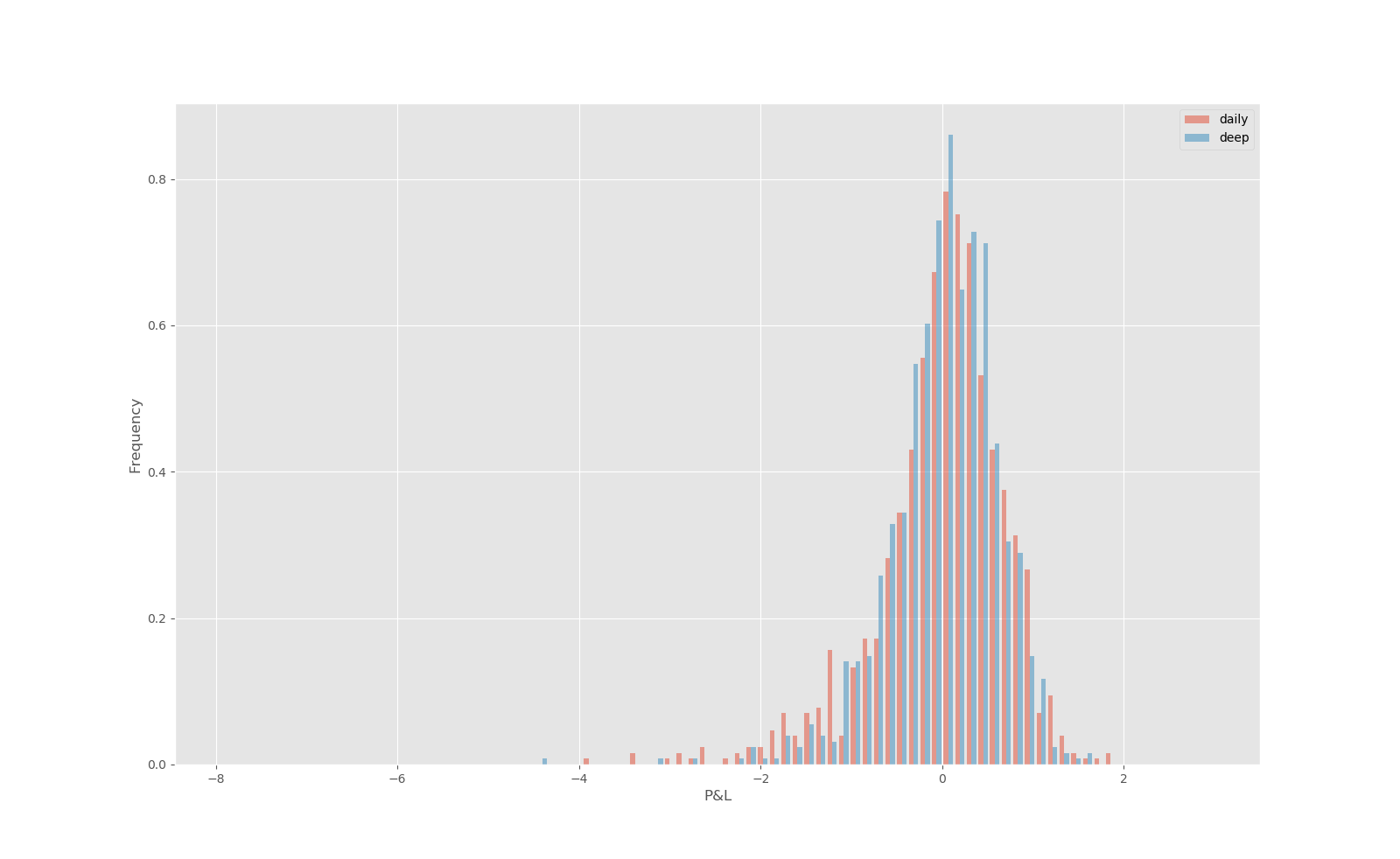}
\caption{$H=0.20$}
\end{subfigure}
\vskip\baselineskip
\begin{subfigure}{.5\textwidth}
\includegraphics[width=1.\linewidth]{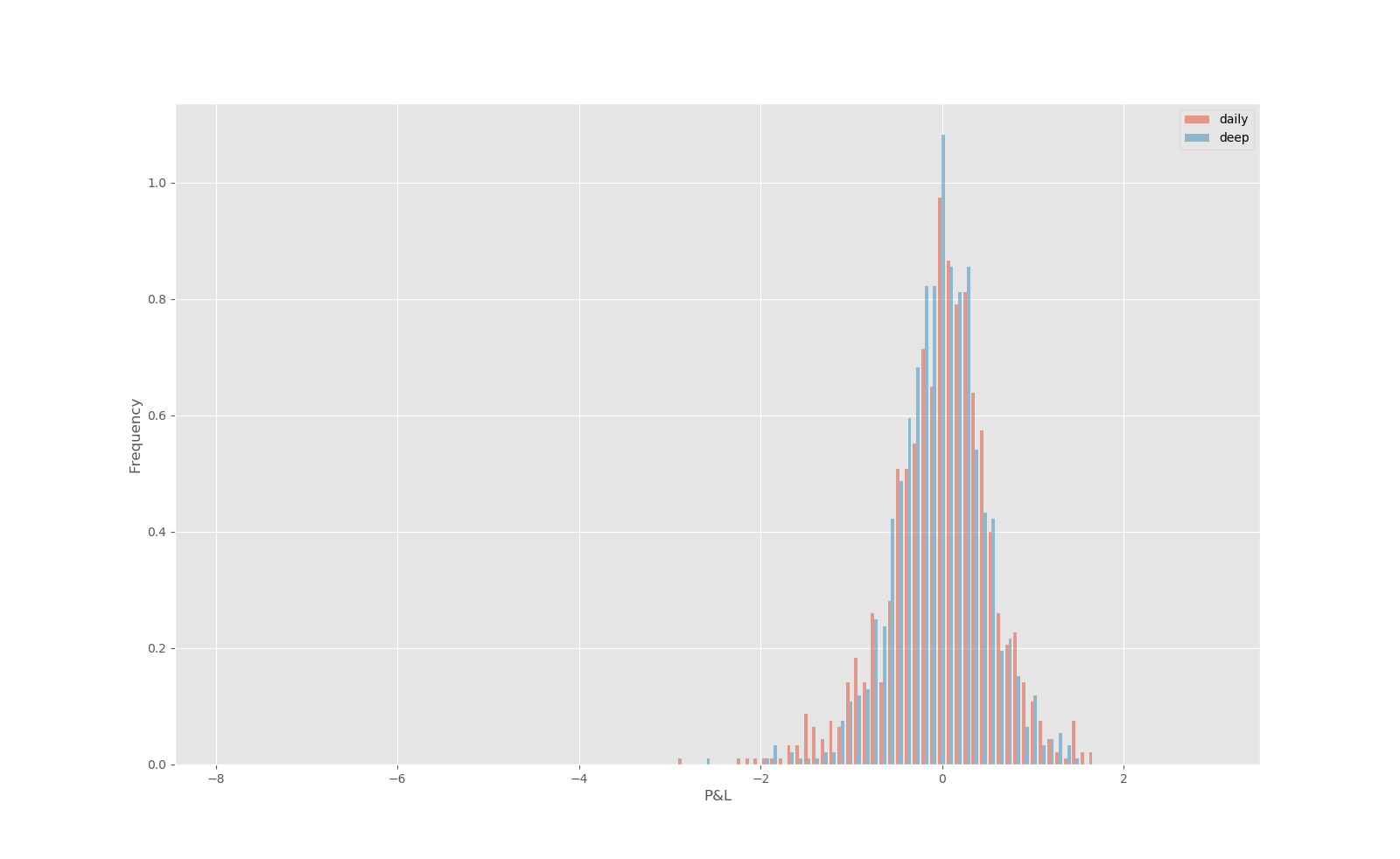}
\caption{$H=0.30$}
\end{subfigure}%
\begin{subfigure}{.5\textwidth}
\includegraphics[width=1.\linewidth]{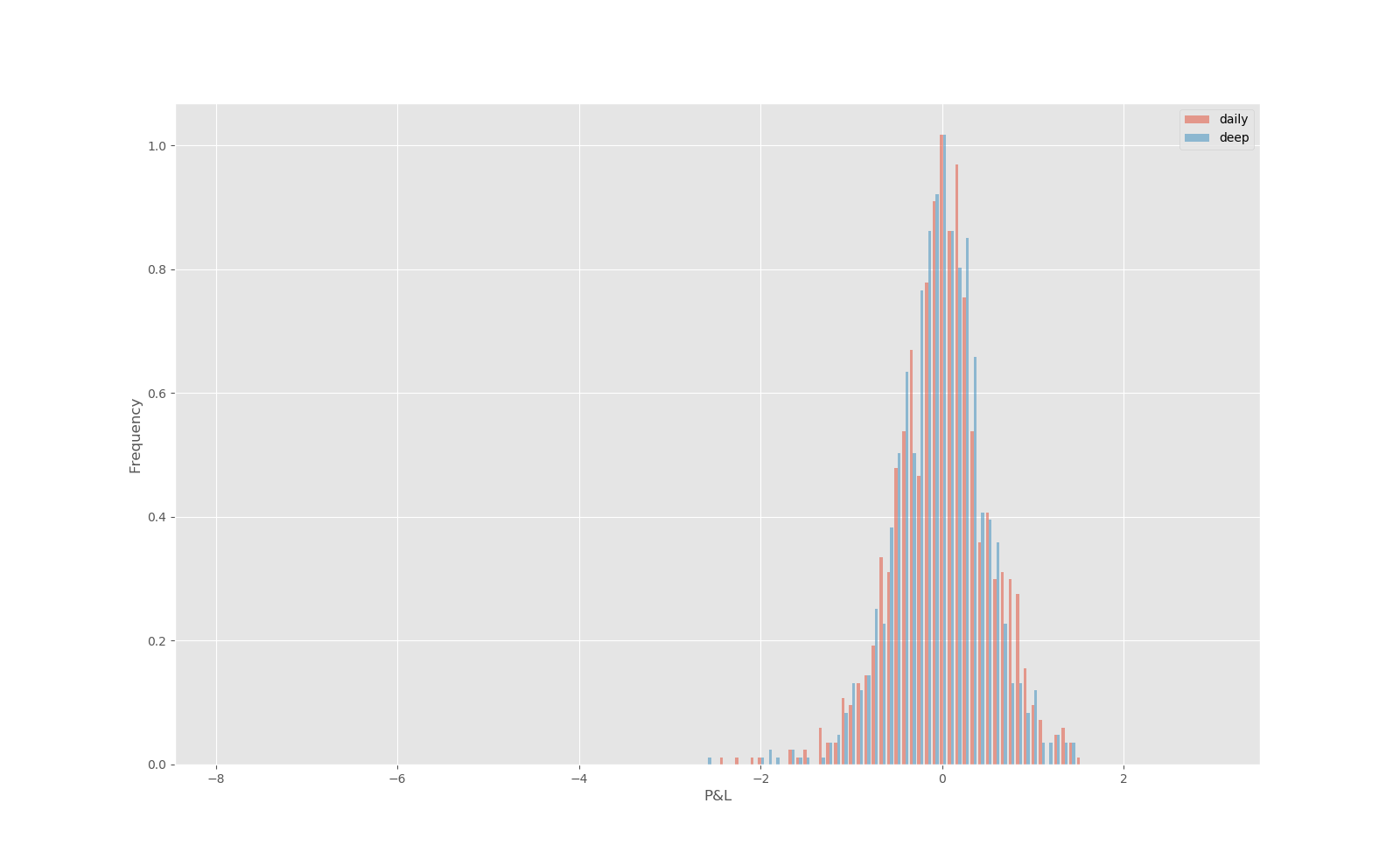}
\caption{$H=0.40$}
\end{subfigure}
\caption{P\&L distributions of rBergomi model hedge (\textit{red}) vs. Deep hedge with proposed architecture (\textit{blue}) for different Hurst parameters realized on $10^3$ sample paths.}\label{fig:bergvsdeepH}
\end{figure}

\newpage
\subsection{Rehedges}

We implement deep hedges on rBergomi paths with the Hurst parameter $H=0.10$, where we re-hedged from every two days upto four times a day. Again, one can see the distribution became slightly less leptocurtic, with more frequent rebalancing. The quadratic losses also decreased with higher frequency. Yet, this seems to happen at a slower rate than expected. This would essentially mean, that as soon as transaction costs are present, small gains from more frequent rebalancing would be completely outweighed by higher transaction
fees. As the matter of fact, for the four-time daily re-hedge the loss slightly increased, which indicates the model once again saturates, this time with respect to the hedging frequency. This is quite surprising considering higher hedging frequency usually translates to better performance in a continuous models. This is because the approximation is getting closer and closer to the continuous setting.

\begin{figure}[hbt!]
\includegraphics[width=1.075\linewidth]{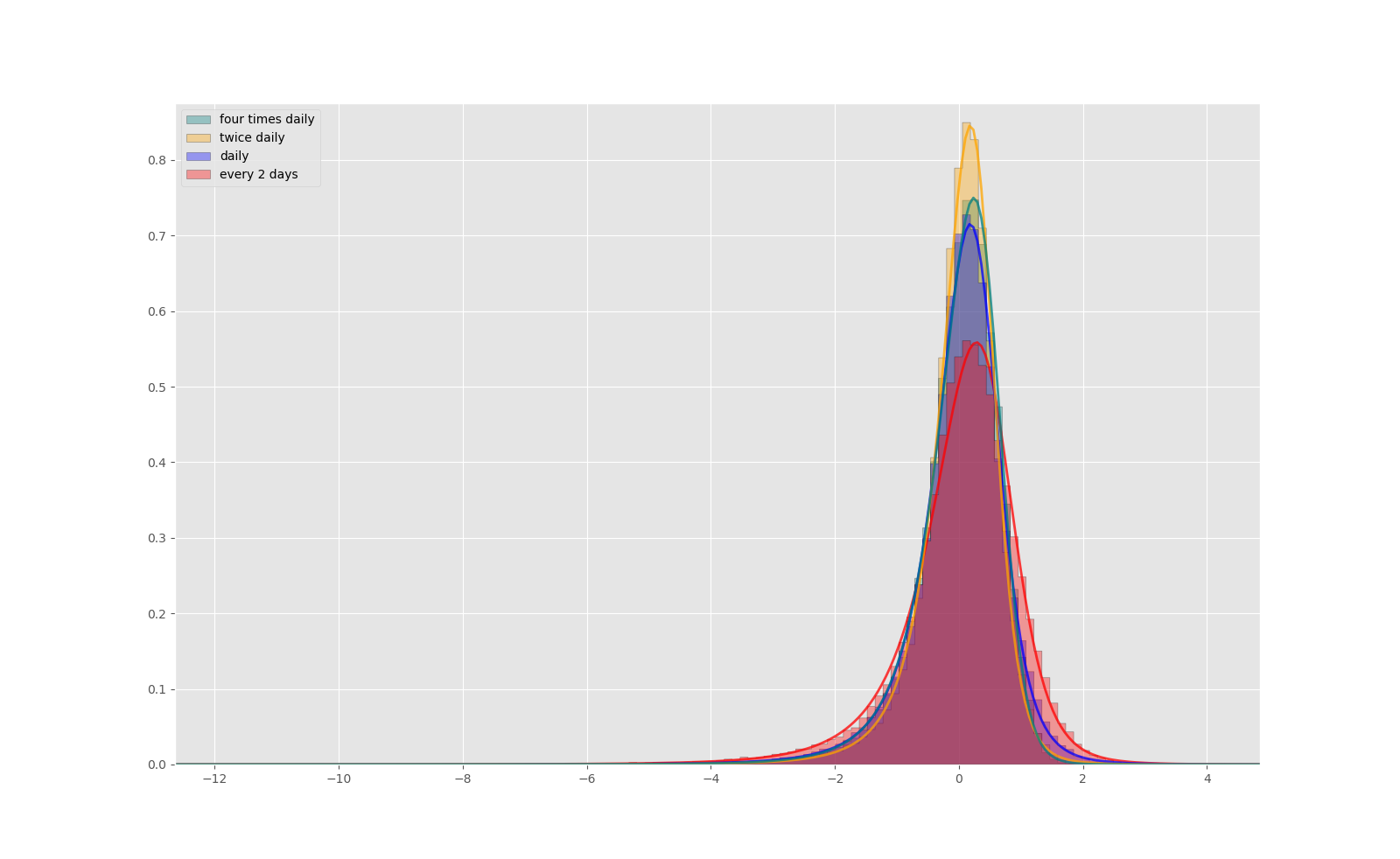}
\caption{Deep hedge for different hedging frequencies (with $H=0.10$). The plot was rescaled to $[-12.5, 4.5]$, because of the massive outliers in the two day rehedge case. Non-central t distribution was fit for better visiblity.}\label{fig:rBergomi01-deep-1-2-4rehedge}
\end{figure}

\begin{table}[H]
\begin{tabular}{ccccc}
\hline
&  \multicolumn{4}{c}{Rehedging frequency} \\
\hline
$H = 0.10$ & Every two days  & Daily & Twice daily  & Four times daily \\ \hline \hline
\multicolumn{1}{c|}{Quadratic loss} & 1.11 & 0.65 & 0.46 & 0.52 \\
\multicolumn{1}{c|}{Training time (h)} & 3.1 & 7.5 & 19.6 & 45.3  
\end{tabular}
\caption{Comparison of the deep hedge quadratic losses for different hedging frequencies (with $H=0.10$).}\label{table:bergvsdeepH}
\end{table}

Behaviour of distributions as well as hedging losses is in fact quite reminiscent of the behaviour of jump diffusion models analysed by A. Sepp \cite{Sepp2012}, which we recall in the following section.

\subsection{Relation to the literature}\label{sec:relationtolit}

It is rather interesting that A. Sepp \cite{Sepp2012} observes a similar behaviour, when delta hedging under jump diffusion models. Similarly to our observations above, he finds (in presence of jumps) that after a certain point the volatility of the P{\&}L cannot be reduced by increasing the hedging frequency. More precisely, he shows that for jump diffusion models, there is a lower bound on the volatility of the P{\&}L in relation to the hedging frequency. Not only that, the P{\&}L distributions in Figure \ref{fig:seppdist} for delta hedges under jump diffusion models are generally fairly similar to ours. 

\begin{figure}[!hbt]
\centering
\includegraphics[width=0.9\linewidth]{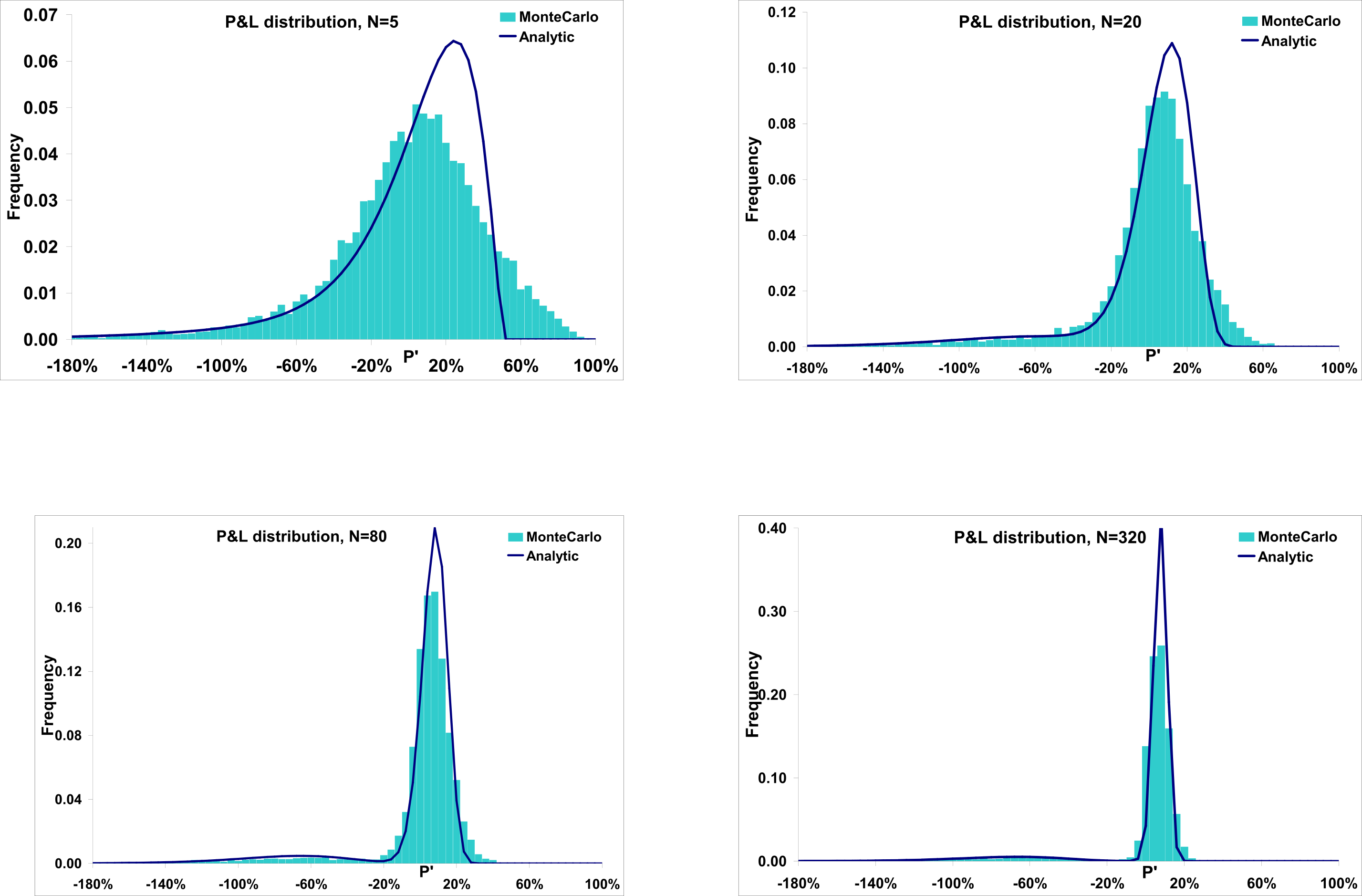}
\caption{P\&L distributions for delta hedging under jump diffusion models (JDM) from \citep{Sepp2012}.}\label{fig:seppdist}
\end{figure}

This gives us the idea to treat the discretsed rough models as jump models. In this case the market is incomplete and it is not possible to perfectly hedge a contingent claim with a portfolio containing a finite number of instruments \citep{He2007}. In practice traders try to come as close as possible to the perfect hedge by trading a number of different options. 

Unfortunately, when trying to implement the hedge approximation, we are quickly faced with the absence of analytical pricing formulas and limitations of the slow Monte-Carlo scheme. In order for us to train the deep hedge, we would have to calculate option prices  on every time step of each sample path. In a typical application we would need around $10$ options with different strikes and at least $10^5$ sample paths.

\section{Conclusion}\label{sec:conclusion}

In this work, we presented and compared different methods for hedging under rough volatility models. More specifically, we analysed and implemented the perfect hedge for the rBergomi model from \citep{Viens2019} and used the deep hedging scheme from \citep{Buehler2019}, which had to be adapted to a non-Markovian framework.

We were particularly interested in the dependence of the P{\&}L on the Hurst parameter. We conclude the deep hedge with the proposed architecture performs better than the discretized perfect hedge for all $H$. We also find that the hedging P{\&}L distributions for low $H$ are highly left-skewed and have a lot of mass in the left tail under the model hedge as well as the deep hedge.

To mitigate the heavy losses in cases when $H$ is close to zero, we explored increasing the hedging frequency upto four times a day. The loss did improve and the P{\&}L distribution became less leptocurtic, however only slightly.

Intriguingly, slow response to increased hedging frequency and left-skewed P{\&}L distribution are characteristic for delta hedges under jump diffusion models \citep{Sepp2012}. We therefore observe that in terms of hedging there is a relation between jump diffusion models and rough models. In accordance with the literature we find that the price process, despite being a continuous martingale, exhibits jump-like behaviour \citep{McCrickerd2018}. We believe this is an excellent illustration of rough volatility models dynamics. Explosive almost jump-like pattern in the stock price might be the reason why they can fit the short end of implied volatility so well.

In our view, it is crucial to take into account the jump aspect, when looking for an optimal hedge in discretized rough volatility models. Our suggestion for future research is adapting the objective function in deep hedge scheme for jump risk optimization. First step would be optimization of the Expected shortfall risk measure. Next, more appropriate jump risk measures for discretized rough models can be developed. These risk measures cannot be completely analogous to the risk measures in \citep{Sepp2012}, since rough models themselves do not feature jumps.

\begin{appendix}
\section{Appendix}

\subsection{Path derivatives}\label{apx:pathderivatives}
Denote by $\mathcal{D}_0$ a c\`{a}dl\`{a}g space and by $\mathcal{D}_t$ and $\mathcal{C}_t$ the space of c\`{a}dl\`{a}g functions on $[t,T]$ and the space of continuous functions on $[t,T]$ respectively. Additionally, we denote by $\omega$ the sample paths on $[0, T]$, $\omega_t$ as its value at time $t$ and define
\begin{align*}
& \Lambda := [0, T]\times \mathcal{C}([0,T], \mathbb{R}^d), & \bar{\Lambda} := \left\lbrace (t, \omega)\in [0, T]\times \mathcal{D}_0 : \omega\vert _{[t, T]} \in \mathcal{C} \right\rbrace ; \\
&\left\lVert\omega\right\rVert_T := \sup_{t\in [0, T]} \lvert \omega_t \rvert, & \mathbf{d} \left( (t, \omega), (t', \omega ') \right) := \lvert t - t' \rvert + \lVert \omega - \omega ' \rVert _T.
\end{align*}
Furthermore, we denote the set of all $\mathbf{d}$-continuous functions $u : \bar{\Lambda} \rightarrow \mathbb{R}$ by $\mathcal{C} (\bar{\Lambda})$. Define the usual horizontal time derivative for $u\in \mathcal{C} (\bar{\Lambda})$ as in \citep{Dupire2019}:
\begin{equation}
\partial_t u(t,\omega) := \lim_{\delta\downarrow 0}\frac{u(t+\delta, \omega)-u(t,\omega)}{\delta} \qquad \text{ for all } (t,\omega)\in\bar{\Lambda},
\end{equation}
requiring of course that the limit exists. For the spatial derivative with respect to $\omega$, however, we use the definition of the Gateaux derivative for any $(t,\omega)\in\bar{\Lambda}$:
\begin{equation}\label{eq:gateuxderivative}
\left\langle \partial_{\omega} u(t, \omega), \eta \right\rangle = \lim_{\varepsilon \rightarrow 0} \frac{u(t, \omega + \varepsilon\eta\mathbbm{1}_{[t,T]}) - u(t,\omega)}{\varepsilon} \qquad \text{for any } \eta\in\mathcal{C}_t.
\end{equation}
Note that the function $u(t,\cdot)$ in the definition of the derivative is ``lifted'' only on $[t, T]$ and not on $[0, t)$. Hence the convention we follow is actually
\[
\left\langle \partial_{\omega} u(t, \omega), \eta \right\rangle : = \left\langle \partial_{\omega} u(t, \omega), \eta\mathbbm{1}_{[t, T]} \right\rangle \qquad \text{for any } s<t \text{ and }  \eta\in\mathcal{C}_s.
\]
The definition of Gateaux derivative is clearly also equal to
\[
\left\langle \partial_{\omega} u(t, \omega), \eta \right\rangle = \frac{d}{d \varepsilon} \left. u (t, \omega + \varepsilon \eta \mathbbm{1}_{[t, T]}) \right\rvert_{\epsilon = 0}.
\]
\begin{Remark}
We remark that our definition of the spatial derivative is different from the one in \citep{Dupire2019, Cont2013}, where functional derivative quantifies the sensitivity of the functional to the variation solely in the end point of the path i.e. $\omega_t$. While in our definition the perturbation takes place throughout the whole interval $[t,T]$.
\end{Remark}
We define two more spaces necessary for our analysis:
\begin{align*}
\mathcal{C}^{1,2} (\bar{\Lambda}) := & \left\{ u \in \mathcal{C}(\bar{\Lambda}): \varphi \in \mathcal{C}(\bar{\Lambda}) \text{ for } \varphi \in \{ \partial_t u, \partial_\omega u, \partial_{\omega \omega}^2 u \} \right\},\\
\mathcal{C}^{1,2}_+ (\bar{\Lambda}) := & \left\{ u \in \mathcal{C}(\bar{\Lambda}): \varphi \text{ has polynomial growth for } \varphi \in \{ \partial_t u, \partial_\omega u, \partial_{\omega \omega}^2 u \} \text{ and }\right. \\ & \left. \left\langle \partial_{\omega\omega}^2 u, (\eta,\eta) \right\rangle \text{ is locally uniformly continuous in } \omega \text{ with polynomial growth} \right\}.
\end{align*}
\pagebreak

\subsection{Functional It\^o formula}

We have to differentiate two cases. The regular case where $H\in(\frac{1}{2}, 1)$ and the singular case where the coefficients $b$, $\sigma$ explode, because the power-kernel in Riemann-Liouville fractional Brownian motion whenever the Hurst exponent $H$ lies in $(0,\frac{1}{2})$. In the singular case the coefficients $b, \sigma \notin \mathcal{C}_t$ and thus they cannot serve as the test function in the right side of \eqref{eq:gateuxderivative}, since Gateaux derivative would not make sense any more. In order to develop an It\^o formula for the singular case, definitions need to be slightly amended. Nonetheless, Viens et al. show that both cases yield similar Functional Itô formula.

\begin{Assumption}\label{ass:technicalito}\leavevmode
\begin{enumerate}[label=\roman*]
\item The SDE \eqref{eq:volterraframework} admits a weak solution $(X, W)$.
\item $\mathbb{E}\left[ \sup_{t\in [0,T]} \left\vert X_t \right\vert ^p \right] < \infty$ for all $p\geq 1$.
\end{enumerate}
\end{Assumption}

\begin{Assumption}\label{ass:funcito}\leavevmode
\begin{enumerate}[label=\roman*]
\item\sloppy (Regular case) For any $r\in[0,T]$, $\partial_t b(t; r, \cdot), \partial_t \sigma (t; r, \cdot)$ exist for $t\in [r, T]$ and for $\varphi = b,\sigma,\partial_t b, \partial_t \sigma$,
\[
\vert \varphi (t; r, \omega)\vert \leq C_0 (1 + \lVert \omega \rVert ^{\kappa_0}_T) \qquad  C_0, \kappa_0 > 0.
\]
\item\sloppy (Singular case) For any $r\in[0,T]$, $\partial_t (t;r, \cdot)$ exists for $t\in(r,T]$ with $\varphi=b, \sigma$. There exists $H\in(0,\frac{1}{2})$ s.t., for some $C_0, \kappa_0 > 0$
\[
\vert \varphi (t; r, \omega)\vert \leq C_0 (1 + \lVert \omega \rVert ^{\kappa_0}_T)(t-r)^{H-\frac{1}{2}} \text{ and } \vert \varphi_t (t; r, \omega)\vert \leq C_0 (1 + \lVert \omega \rVert ^{\kappa_0}_T)(t-r)^{H-\frac{3}{2}}
\]
\end{enumerate}
\end{Assumption}

\begin{Theorem}[Functional It{\^o} formula]
Let $X$ be a weak solution to the SDE \eqref{eq:volterraframework} for which $\mathbb{E}\left[ \sup_{t\in [0,T]} \left\vert X_t \right\vert ^p \right] < \infty$ for all $p\geq 1$ and Assumption \ref{ass:funcito} hold. Then
\begin{align}\label{eq:funcitoformula}
\begin{split}
d u (t, X\otimes_t\Theta^t) = \partial_t u (t, X\otimes_t\Theta^t)d t + \frac{1}{2}\left\langle \partial_{\omega\omega} u(t, X\otimes_t\Theta^t), (\sigma^{t, X}, \sigma^{t, X}) \right\rangle d t + \\ \left\langle \partial_{\omega} u (t, X\otimes_t\Theta^t), b^{t, X} \right\rangle d t + \left\langle \partial_{\omega} u(t, X\otimes_t\Theta^t), \sigma^{t, X} \right\rangle d W_t, \quad \mathbb{P}\text{-a.s.}
\end{split}
\end{align}
for $u\in C_+^{1,2}(\Lambda)$ in the regular case and $u\in C_{+,\beta}^{1,2}(\Lambda)$ with regularized Gateaux derivative for the singular case. For $\varphi = b, \sigma$ the notation $\varphi_s^{t, \omega} := \varphi(s; t, \omega)$ only emphasizes the dependence on $s\in[t,T]$. For the definition of $C_{+,\beta}^{1,2}(\Lambda)$ and precise statement of the theorem in the singular case see Theorem 3.17 and Theorem 3.10 in \citep{Viens2019}
\end{Theorem}
\pagebreak

\subsection{Discretization of the Gateaux Derivative}\label{apx:discretegateux}

It can be easily shown that $\hat{\Theta}_s^t = f(\Theta_s^t)$ for some $f:\mathbb{R}\rightarrow\mathbb{R}$. Therefore, we have direct relation between the auxiliary process $\Theta$ and the forward variance $\hat{\Theta}$, which allows us to write the option price as the function of the entire \emph{forward variance curve} $\smash{\hat{\Theta}_{[t,T]}^t}$ at time $t\in[0,T]$, namely $\smash{u(t,S_t,\Theta_{[t,T]}^t) = \tilde{u}(t,S_t, \hat{\Theta}_{[t,T]}^t)}$. This is important, when performing Monte-Carlo, since in the rough Bergomi model, the forward variance curve is directly modelled in the variance process with $\xi_t(\cdot) = \hat{\Theta}_{\cdot}^t$.\\
Let us suppose that we are able to trade at times $0=t_0<t_1<\dots<t_n = T$. In order to get the hedging weights at trading times $t_i$, we have to discretize the derivatives. The Gateaux derivative with respect to the stock simplifies to the usual derivative and the discretization is straightforward:
\begin{align}
\partial_x \tilde{u}(t, S_t, \hat{\Theta}_{[t,T]}^t) \approx \frac{\tilde{u}(t, S_t + \varepsilon, \hat{\Theta}_{[t, T]}^t) - \tilde{u}(t, S_t, \hat{\Theta}_{[t, T]}^t)}{\varepsilon} \qquad \text{for small } \varepsilon > 0.
\end{align}
For the path-wise derivative the discretization is not immediately obvious, especially because of the dependence of the option price $\tilde{u}$ at time $t$ on functional over the whole interval $[t,T]$, more precisely since $u:[0,T]\times[0,\infty)\times\mathcal{C}([0,T]\rightarrow \mathbb{R})$. First, we remind ourselves of the definition of the Gateaux derivative on a path $\omega$:
\[
\left\langle \partial_{\omega} u(t, \omega), \eta \right\rangle = \lim_{\varepsilon \rightarrow 0} \frac{u(t, \omega + \varepsilon\eta\mathbbm{1}_{[t,T]}) - u(t,\omega)}{\varepsilon} \qquad \text{for any } \eta\in\Omega_t.
\]
We proceed as in \citep{Jacquier2019} by approximating $\hat{\Theta}_{[t,T]}^t$ as a piecewise constant function
\begin{align}
&\hat{\Theta}_s^t \approx \sum_{i\in\mathcal{I}} \hat{\Theta}_i^t \mathbbm{1}_{[t_i, t_{i+1})}(s)
&a^t_i = a^t_s \mathbbm{1}_{[t_i, t_{i+1})}(s),
\end{align}
where $\mathcal{I}:= \{i\in \mathbb{N}: t\leq t_i \leq T\}$. We introduce the following approximations of the path derivatives along the direction $a^t$:
\begin{align*}
\left\langle \partial_\omega \tilde{u} (t, S_t, \hat{\Theta}_{[t,T]}^t), a^t \right\rangle &\approx \left. \partial_\varepsilon \tilde{u}\left(t, S_t, \sum_{i\in\mathcal{I}} \left.(\hat{\Theta}_i^t + \varepsilon a^t)\mathbbm{1}_{[t_i, t_{i+1})}(s)\right\vert_{s\in[t,T]}\right)\right\vert_{\varepsilon=0}\\
&= \left.\partial_\varepsilon\hat{u}\left(t, S_t, \left(\hat{\Theta}_i^t + \varepsilon a^t_i\right)_{i\in\mathcal{I}}\right)\right\vert_{\varepsilon=0}\\
&= \sum_{i\in\mathcal{I}} \partial_{\hat{\Theta}_i^t} \hat{u}(t, S_t, \boldsymbol{\theta}^t)a_i^t,
\end{align*}
with $\boldsymbol{\theta}^t := (\hat{\Theta}^t_i)_{i\in\mathcal{I}}$ and $\hat{u}$ acts on $[0,T]\times[0,\infty)\times\mathbb{R}^{\#\mathcal{I}}$. Further discretizing the derivative we have for the flat forward variance $\xi(t) = \xi_0$:
\[
\left\langle \partial_\omega \tilde{u} (t, S_t, \xi_{0}), a^t \right\rangle \approx \frac{\tilde{u}(t, S_t, \xi_0 + \varepsilon) - \tilde{u}(t, S_t, \xi_0)}{\varepsilon} a^t \qquad \text{for small } \varepsilon > 0.
\]
The option prices $\tilde{u}$ can now be evaluated using Monte-Carlo at each time step to get the hedging weights. Note that the discretization of the Gateaux derivative is purely heuristic and that a rigorous proof of the convergence to the true derivative is out of scope of this work. For more details we refer to \citep{Jacquier2019}.

\end{appendix}


\begin{thebibliography}{10}

\bibitem{Hernandez2016}
A.~Hernandez, Model Calibration with Neural Networks
  \href{http://dx.doi.org/10.2139/ssrn.2812140}{{\em Risk} (2016) }.

\bibitem{Horvath2019}
B.~Horvath, A.~Muguruza, and M.~Tomas, Deep Learning Volatility
  \href{http://arxiv.org/abs/1901.09647}{{\ttfamily arXiv:1901.09647
  [q-fin.MF]}}.

\bibitem{Bayer2019}
C.~Bayer, B.~Horvath, A.~Muguruza, B.~Stemper, and M.~Tomas, On deep
  calibration of (rough) stochastic volatility models
  \href{http://arxiv.org/abs/1908.08806}{{\ttfamily arXiv:1908.08806
  [q-fin.MF]}}.

\bibitem{Liu2019}
S.~Liu, A.~Borovykh, L.~A. Grzelak, and C.~W. Oosterlee, A neural network-based
  framework for financial model calibration {\em Journal of Mathematics in
  Industry} {\bfseries 9} no.~1, (2019) 9.

\bibitem{Ruf2019}
J.~Ruf and W.~Wang, Neural networks for option pricing and hedging: a
  literature review {\em Available at SSRN 3486363} (2019) .

\bibitem{Benth2020}
F.~E. Benth, N.~Detering, and S.~Lavagnini, Accuracy of Deep Learning in
  Calibrating {HJM} Forward Curves {\em arXiv preprint arXiv:2006.01911} (2020)
  .

\bibitem{Cuchiero2020}
C.~Cuchiero, W.~Khosrawi, and J.~Teichmann, A Generative Adversarial Network
  Approach to Calibration of Local Stochastic Volatility Models
  \href{http://dx.doi.org/10.3390/risks8040101}{{\em Risks} {\bfseries 8}
  no.~4, (Sep, 2020) 101}.

\bibitem{Gierjatowicz2020}
P.~Gierjatowicz, M.~Sabate-Vidales, D.~Siska, L.~Szpruch, and Z.~Zuric, Robust
  Pricing and Hedging via Neural {SDEs}
  \href{http://dx.doi.org/10.2139/ssrn.3646241}{{\em {SSRN} Electronic Journal}
  (2020) }.

\bibitem{Buehler2019}
H.~Buehler, L.~Gonon, J.~Teichmann, and B.~Wood, Deep hedging
  \href{http://dx.doi.org/10.1080/14697688.2019.1571683}{{\em Quantitative
  Finance} {\bfseries 19} no.~8, (Feb, 2019) 1271--1291}.

\bibitem{HenryLabordere2019}
P.~Henry-Labordere, Generative Models for Financial Data
  \href{http://dx.doi.org/10.2139/ssrn.3408007}{{\em {SSRN} Electronic Journal}
  (2019) }.

\bibitem{Wiese2019}
M.~Wiese, L.~Bai, B.~Wood, and H.~Buehler, Deep Hedging: Learning to Simulate
  Equity Option Markets {\em NeurIPS 2019 Workshop on Robust AI in Financial
  Services: Data, Fairness, Explainability, Trustworthiness, and Privacy}
  (Nov., 2019) , \href{http://arxiv.org/abs/1911.01700}{{\ttfamily
  arXiv:1911.01700 [q-fin.CP]}}.

\bibitem{Wiese2020}
M.~Wiese, R.~Knobloch, R.~Korn, and P.~Kretschmer, Quant {GANs}: deep
  generation of financial time series
  \href{http://dx.doi.org/10.1080/14697688.2020.1730426}{{\em Quantitative
  Finance} {\bfseries 20} no.~9, (Apr, 2020) 1419--1440}.

\bibitem{Kondratyev2020}
C.~S. Alexei~Kondratyev, The Market Generator {\em Risk} (2020) .

\bibitem{Buehler2020}
H.~Buehler, B.~Horvath, T.~Lyons, I.~P. Arribas, and B.~Wood, Generating
  Financial Markets With Signatures
  \href{http://dx.doi.org/10.2139/ssrn.3657366}{{\em {SSRN} Electronic Journal}
  (2020) }.

\bibitem{Buehler2020a}
H.~Buehler, B.~Horvath, T.~Lyons, I.~P. Arribas, and B.~Wood, A Data-Driven
  Market Simulator for Small Data Environments
  \href{http://dx.doi.org/10.2139/ssrn.3632431}{{\em {SSRN} Electronic Journal}
  (2020) }.

\bibitem{Cuchiero2020a}
C.~Cuchiero, M.~Larsson, and J.~Teichmann, Deep Neural Networks, Generic
  Universal Interpolation, and Controlled {ODEs}
  \href{http://dx.doi.org/10.1137/19m1284117}{{\em {SIAM} Journal on
  Mathematics of Data Science} {\bfseries 2} no.~3, (Jan, 2020) 901--919}.

\bibitem{Xu2020}
T.~Xu, L.~K. Wenliang, M.~Munn, and B.~Acciaio, COT-GAN: Generating Sequential
  Data via Causal Optimal Transport
  \href{http://arxiv.org/abs/2006.08571}{{\ttfamily arXiv:2006.08571
  [stat.ML]}}.

\bibitem{Alos2007}
E.~Al{\`o}s, J.~A. Le{\'o}n, and J.~Vives, On the short-time behavior of the
  implied volatility for jump-diffusion models with stochastic volatility {\em
  Finance and Stochastics} {\bfseries 11} no.~4, (2007) 571--589.

\bibitem{Fukasawa2010}
M.~Fukasawa, Asymptotic analysis for stochastic volatility: martingale
  expansion \href{http://dx.doi.org/10.1007/s00780-010-0136-6}{{\em Finance and
  Stochastics} {\bfseries 15} no.~4, (Aug, 2010) 635--654}.

\bibitem{Gatheral2018}
J.~Gatheral, T.~Jaisson, and M.~Rosenbaum, Volatility is rough
  \href{http://dx.doi.org/10.1080/14697688.2017.1393551}{{\em Quantitative
  Finance} {\bfseries 18} no.~6, (Mar, 2018) 933--949}.

\bibitem{Bolko2020}
A.~E. Bolko, K.~Christensen, M.~S. Pakkanen, and B.~Veliyev, Roughness in spot
  variance? A GMM approach for estimation of fractional log-normal stochastic
  volatility models using realized measures
  \href{http://arxiv.org/abs/2010.04610}{{\ttfamily arXiv:2010.04610
  [q-fin.ST]}}.

\bibitem{Livieri2018}
G.~Livieri, S.~Mouti, A.~Pallavicini, and M.~Rosenbaum, Rough volatility:
  Evidence from option prices
  \href{http://dx.doi.org/10.1080/24725854.2018.1444297}{{\em {IISE}
  Transactions} {\bfseries 50} no.~9, (Jun, 2018) 767--776}.

\bibitem{Xu2005}
M.~Xu, Risk measure pricing and hedging in incomplete markets
  \href{http://dx.doi.org/10.1007/s10436-005-0023-x}{{\em Annals of Finance}
  {\bfseries 2} no.~1, (Oct, 2005) 51--71}.

\bibitem{IAR09}
A.~Ilhan, M.~Jonsson, and R.~Sircar, Optimal static-dynamic hedges for exotic
  options under convex risk measures {\em Stochastic Processes and their
  Applications} {\bfseries 119} no.~10, (2009) 3608--3632.

\bibitem{FS16}
F.~Hans and A.~Schied, {\em Stochastic Finance: An Introduction in Discrete
  Time}.
\newblock De Gruyter, 2016.

\bibitem{Bayer2015}
C.~Bayer, P.~Friz, and J.~Gatheral, Pricing under rough volatility
  \href{http://dx.doi.org/10.1080/14697688.2015.1099717}{{\em Quantitative
  Finance} {\bfseries 16} no.~6, (Nov, 2015) 887--904}.

\bibitem{Kac1949}
M.~Kac, On distributions of certain Wiener functionals
  \href{http://dx.doi.org/10.1090/s0002-9947-1949-0027960-x}{{\em Transactions
  of the American Mathematical Society} {\bfseries 65} no.~1, (Jan, 1949)
  1--1}.

\bibitem{Dupire2019}
B.~Dupire, Functional It{\^{o}} calculus
  \href{http://dx.doi.org/10.1080/14697688.2019.1575974}{{\em Quantitative
  Finance} {\bfseries 19} no.~5, (Apr, 2019) 721--729}.

\bibitem{Cont2013}
R.~Cont and D.-A. Fourni{\'{e}}, Functional It{\^{o}} calculus and stochastic
  integral representation of martingales
  \href{http://dx.doi.org/10.1214/11-aop721}{{\em The Annals of Probability}
  {\bfseries 41} no.~1, (Jan, 2013) 109--133}.

\bibitem{Viens2019}
F.~Viens and J.~Zhang, A martingale approach for fractional Brownian motions
  and related path dependent {PDEs}
  \href{http://dx.doi.org/10.1214/19-aap1486}{{\em The Annals of Applied
  Probability} {\bfseries 29} no.~6, (Dec, 2019) 3489--3540}.

\bibitem{Bennedsen2017}
M.~Bennedsen, A.~Lunde, and M.~S. Pakkanen, Hybrid scheme for Brownian
  semistationary processes
  \href{http://dx.doi.org/10.1007/s00780-017-0335-5}{{\em Finance and
  Stochastics} {\bfseries 21} no.~4, (Jun, 2017) 931--965}.

\bibitem{McCrickerd2018}
R.~McCrickerd and M.~S. Pakkanen, Turbocharging Monte Carlo pricing for the
  rough Bergomi model
  \href{http://dx.doi.org/10.1080/14697688.2018.1459812}{{\em Quantitative
  Finance} {\bfseries 18} no.~11, (Apr, 2018) 1877--1886}.

\bibitem{Carmona1998}
P.~Carmona, G.~Montseny, and L.~Coutin, Application of a representation of long
  memory gaussian processes {\em Publications du Laboratoire de statistique et
  probabilit{\'e}s} (1998) .

\bibitem{Matteo2005}
T.~Di~Matteo, T.~Aste, and M.~M. Dacorogna, Long-term memories of developed and
  emerging markets: Using the scaling analysis to characterize their stage of
  development {\em Journal of Banking \& Finance} {\bfseries 29} no.~4, (2005)
  827--851.

\bibitem{Matteo2007}
T.~Di~Matteo, Multi-scaling in finance {\em Quantitative finance} {\bfseries 7}
  no.~1, (2007) 21--36.

\bibitem{Pascanu2013}
K.~C. Y.~B. Razvan~Pascanu, Caglar~Gulcehre, How to construct deep recurrent
  neural networks in {\em Proceedings of the Second International Conference on
  Learning Representations}.
\newblock 2014.

\bibitem{Hochreiter1997}
S.~Hochreiter and J.~Schmidhuber, Long Short-Term Memory
  \href{http://dx.doi.org/10.1162/neco.1997.9.8.1735}{{\em Neural Computation}
  {\bfseries 9} no.~8, (Nov, 1997) 1735--1780}.

\bibitem{Gassiat2019}
P.~Gassiat, On the martingale property in the rough Bergomi model
  \href{http://dx.doi.org/10.1214/19-ecp239}{{\em Electronic Communications in
  Probability} {\bfseries 24} no.~0, (2019) }.

\bibitem{Sepp2012}
A.~Sepp, An approximate distribution of delta-hedging errors in a
  jump-diffusion model with discrete trading and transaction costs
  \href{http://dx.doi.org/10.1080/14697688.2010.494613}{{\em Quantitative
  Finance} {\bfseries 12} no.~7, (Jul, 2012) 1119--1141}.

\bibitem{He2007}
C.~He, J.~S. Kennedy, T.~F. Coleman, P.~A. Forsyth, Y.~Li, and K.~R. Vetzal,
  Calibration and hedging under jump diffusion
  \href{http://dx.doi.org/10.1007/s11147-006-9003-1}{{\em Review of Derivatives
  Research} {\bfseries 9} no.~1, (Jan, 2007) 1--35}.

\bibitem{Jacquier2019}
A.~Jacquier and M.~Oumgari, Deep Curve-dependent PDEs for affine rough
  volatility \href{http://arxiv.org/abs/1906.02551}{{\ttfamily arXiv:1906.02551
  [q-fin.PR]}}.

\end{thebibliography}

\providecommand{\href}[2]{#2}\begingroup\raggedright\endgroup

\end{document}